\documentclass{article}
\usepackage{ijcai24}

\usepackage[table,xcdraw]{xcolor}
\usepackage{times}
\usepackage{soul}
\usepackage{url}
\usepackage[hidelinks]{hyperref}
\usepackage[utf8]{inputenc}
\usepackage[small]{caption}
\usepackage{graphicx}
\usepackage{amsmath}
\usepackage{amsthm}
\usepackage{booktabs}
\usepackage[ruled,vlined,linesnumbered]{algorithm2e}      
\usepackage[switch]{lineno}

\usepackage{graphicx}%插入图
\usepackage{subfigure}
\usepackage{multirow}
\usepackage{booktabs}

\usepackage[utf8]{inputenc}
\usepackage{url}
\usepackage{booktabs}
\usepackage{amssymb}
\usepackage{bbding}
\usepackage{pifont}
\usepackage{wasysym}
\usepackage{utfsym}
\usepackage{fontawesome}
\usepackage{color}
\usepackage{eso-pic}
\usepackage[marginal]{footmisc}

\renewcommand{\thefootnote}{}

\def\ie{\textit{i.e.}}
 
\def\eg{\textit{e.g.}}

\title{DarkFed: A Data-Free Backdoor Attack in Federated Learning}

\author{
    Author Name
    \affiliations
    Affiliation
    \emails
    email@example.com
}

\author{
Minghui Li$^{1}$\footnotemark[1]\and
Wei Wan$^{2,3,4,5,6}$\footnotemark[1]\footnotemark[2]\and
Yuxuan Ning$^{7}$\footnotemark[1]\and
Shengshan Hu$^{2,3,4,5,6}$\and\\
Lulu Xue$^{2,3,4,5,6}$\and
Leo Yu Zhang$^8$\and
Yichen Wang$^{2,3,4,5,6}$\
\affiliations
$^1$School of Software Engineering, Huazhong University of Science and Technology\\
$^2$National Engineering Research Center for Big Data Technology and System\\
$^3$Services Computing Technology and System Lab  \\
$^4$Hubei Engineering Research Center on Big Data Security\\
$^5$Hubei Key Laboratory of Distributed System Security\\
$^6$School of Cyber Science and Engineering, Huazhong University of Science and Technology\\
$^7$School of Computer Science and Technology, Huazhong University of Science and Technology\\
$^8$School of Information and Communication Technology, Griffith University
\emails
\{minghuili, wanwei\_0303, ningyuxuan, hushengshan, lluxue, wangyichen\}@hust.edu.cn,\\
leo.zhang@griffith.edu.au
}

\begin{document}
\AddToShipoutPictureBG*{
  \AtPageUpperLeft{
    \put(0,-20){\makebox[\paperwidth][c]{\color{blue}\textbf{This paper has been accepted by IJCAI 2024}}}
  }
}

\maketitle
\renewcommand{\thefootnote}{\fnsymbol{footnote}}
\footnotetext[1]{These authors contributed equally to this work.}
\footnotetext[2]{Corresponding author.}
\begin{abstract}
\textit{Federated learning} (FL) has been demonstrated to be susceptible to backdoor attacks. However, existing academic studies on FL backdoor attacks rely on a high proportion of real clients with main task-related data, which is impractical. In the context of real-world industrial scenarios, even the simplest defense suffices to defend against the state-of-the-art attack, 3DFed. A practical FL backdoor attack remains in a nascent stage of development.

To bridge this gap, we present DarkFed. Initially, we emulate a series of fake clients, thereby achieving the attacker proportion typical of academic research scenarios. Given that these emulated fake clients lack genuine training data, we further propose a data-free approach to backdoor FL. 
Specifically, we delve into the feasibility of injecting a backdoor using a shadow dataset. Our exploration reveals that impressive attack performance can be achieved, even when there is a substantial gap between the shadow dataset and the main task dataset. This holds true even when employing synthetic data devoid of any semantic information as the shadow dataset.
Subsequently, we strategically construct a series of covert backdoor updates in an optimized manner, mimicking the properties of benign updates, to evade detection by defenses. A substantial body of empirical evidence validates the tangible effectiveness of DarkFed.
\end{abstract}

\section{Introduction}
\textit{Federated learning} (FL)~\cite{FedAvg,lu2023preserving}, one of the prevailing distributed paradigms, facilitates the collaborative construction of a high-precision global model by multiple clients with small amounts of data, all under the coordination of a central server. Notably, FL excels at preserving privacy since clients' training data remains localized throughout the entire model construction process.

However, the distributed nature of FL also presents a significant challenge: the central server struggles to discern the quality of client-uploaded parameters. Consequently, FL faces a severe threat known as poison attacks~\cite{FLSurvey,lu2024depriving}. These attacks can be categorized into two main types: Byzantine attacks~\cite{ZHT-IJCAI,misa}, and backdoor attacks~\cite{zhang2024detector,zhang2024stealthy} The former aims to reduce the global model's recognition accuracy for all samples, while the latter specifically misclassifies samples specified by the adversary without affecting the model's recognition of normal samples. This indicates that backdoor attacks are more covert and insidious compared to Byzantine attacks.
% Imagine an autonomous vehicle deploying a traffic sign recognition model trained using FL (shown in Fig.~\ref{fig:BackdoorExample}). If this model is injected with a backdoor, it consistently recognizes regular traffic signs correctly. However, when encountering samples specified by the attacker (\eg, signs with triggers), the model classifies them as desired by the attacker. For instance, it may recognize a `slow' sign as `speed limit 120 Km/h', a highly dangerous scenario.
Therefore, this paper focuses on backdoor attacks in FL.

% \begin{figure}[t]
% \centerline{\includegraphics[width=1\columnwidth]{BackdoorExample.pdf}}
% \caption{Backdoor attacks in federated leraning}
% \label{fig:BackdoorExample}
% \end{figure}

\begin{table}[!t]
\renewcommand\arraystretch{1}
\centering
%   \vspace{-1.5mm}
\caption{Performance of existing backdoor attacks in academic research scenarios and real-world industrial scenarios.}
  \vspace{-2mm}
\resizebox{\linewidth}{!}{
\begin{tabular}{|c||cc|cc|}
\hline
\multirow{2}{*}{\textbf{Attacks}} & \multicolumn{2}{c|}{\textbf{20\% Attackers}} & \multicolumn{2}{c|}{\textbf{1\% Attackers}} \\ \cline{2-5} 
 & \multicolumn{1}{c|}{\textbf{ACC (\%)}} & \textbf{ASR (\%)} & \multicolumn{1}{c|}{\textbf{ACC (\%)}} & \textbf{ASR (\%)} \\ \hline
Model Replacement & \multicolumn{1}{c|}{90.07} & 97.93 & \multicolumn{1}{c|}{90.64} & 0.53 \\ \hline
3DFed & \multicolumn{1}{c|}{90.14} & 98.71 & \multicolumn{1}{c|}{90.36} & 0.52 \\ \hline
\end{tabular}
}
\vspace{-5mm}
\label{tab: AcademicVSIndustry}
\end{table}

FL is shown to be susceptible to backdoor attacks~\cite{DBA,CerP,3DFed}. However, the success of these attacks critically hinges on a high proportion of genuine attackers possessing samples relevant to the main task. Typically, they require $20\%$ of attackers with authentic training data to successfully inject a backdoor. In real-world industrial scenarios~\cite{BackToTheDrawingBoard}, attackers often constitute only $1\%$ or even less of the total clients. As shown in Tab.~\ref{tab: AcademicVSIndustry}, for both the classical Model Replacement Attack~\cite{HowToBackdoor} and the recent 3DFed~\cite{3DFed}, we consider scenarios with $20\%$ attackers (academic research scenarios) and $1\%$ attackers (real-world industrial scenarios). Notably, we employ only the most primitive defense method, Norm Clipping~\cite{NormClipping}, which restricts the magnitude of local updates to remain within a specified threshold. We observe that in academic research scenarios, these attacks can indeed achieve significant \textit{attack success rate} (ASR) and \textit{accuracy of the model} (ACC). However, surprisingly, in real-world industrial scenarios, even the \textit{state-of-the-art} (SOTA) 3DFed fails to backdoor FL equipped with the simplest defense. We speculate that this is due to the low proportion of attackers, which results in the backdoor task-related knowledge being overshadowed by the main task-related knowledge in the aggregation stage. The result suggests that existing backdoor attacks in FL are impractical, and an effective FL backdoor attack for real-world industrial scenarios is yet to be developed.

In light of this, we embark on the initial steps toward developing backdoor attacks in FL tailored for real-world industrial contexts. Building upon the research in \cite{FakeClients}, we can emulate a series of fake clients using open-source projects or Android emulators. These approaches can significantly increase the number of attackers to match the settings of academic research scenarios. However, these emulated fake clients are unable to provide authentic main task-related data. Consequently, the primary challenge pivots towards devising a data-free backdoor attack in FL.

In this paper, we propose DarkFed, the first \textbf{DA}ta-f\textbf{R}ee bac\textbf{K}door attack in \textbf{FED}erated learning. Specifically, we first explore the impact of shadow datasets on backdoor attacks. Surprisingly, even when there is a substantial gap between the shadow dataset and the main task dataset (\eg, between CIFAR-10 and GTSRB), the backdoor can be successfully implanted while maintaining model utility. What's even more astonishing is that using synthetic data devoid of any semantic information (\eg, generated through a Gaussian distribution) as the shadow dataset still yields significant success in backdoor attacks. These promising results inspire us to inject the backdoor using a shadow dataset on the emulated fake clients. However, directly transferring the previous process is prone to detection by existing defenses due to the significant differences between backdoor updates and benign updates, leading to the failure of the attack. To further enhance the stealthiness of the attack, we propose property mimicry, optimizing backdoor updates to mimic benign updates in terms of magnitude, distribution, and consistency. These properties are widely employed by FL backdoor defenses to detect backdoor updates. This optimization significantly boosts the covert nature of the attack.

In summary, our contributions are as follows:

\begin{itemize}
    \item We introduce DarkFed, the first data-free backdoor attack in FL. This attack does not rely on task-specific data, enabling its use in scenarios with emulated fake clients, thus achieving a practical backdoor attack.
    
    \item We investigate the feasibility of injecting a backdoor with shadow datasets and find that even with synthetic datasets, successful backdoor injection is achievable. We extend this concept into the realm of FL.
    
    % Specifically, the \textit{reliable client selection} reduces the selected probabilities of those previously poor-performed clients so that only a limited number of attackers have a chance to poison the global model. And the \textit{update denoising} can rectify the direction of the slightly noised but harmful updates.
    
    % \item Unlike existing defenses, which are effective in either collusion attack scenario or dispersion attack scenario, the proposed FPD can deal with both scenarios readily.

    \item We introduce a novel defense evasion technique, property mimicry, which enables backdoor updates to mimic the properties of benign updates, thereby enhancing the stealthiness of the attack.

    \item Extensive experiments demonstrate that DarkFed achieves attack effects comparable to SOTA data-dependent attacks.
    
\end{itemize}

\section{Related Work}
\subsection{Backdoor Attacks in FL}
A plethora of studies have suggested that FL is exceptionally susceptible to backdoor attacks~\cite{TED,CRC,PointCRT,Badhash}. \cite{HowToBackdoor} is among the pioneers in launching backdoor attacks on FL. They introduce the Model Replacement Attack, amplifying the magnitude of backdoor updates proportionally, ensuring the dominance of backdoor parameters in the global model. Furthermore, they introduce the Semantic Backdoor Attack, which doesn't require any modifications to the training samples but leverages samples with specific semantic information to trigger the backdoor. For example, this attack classifies all green cars as horses. Drawing inspiration from the Semantic Backdoor Attack, \cite{NormClipping} introduce an edge-case backdoor attack, which uses rare samples (the tail of a dataset) to trigger the backdoor. \cite{DBA} propose DBA (\textit{Distributed Backdoor Attack}), which decomposes a trigger into multiple sub-triggers, with each attacker holding one of these sub-triggers for data poisoning. Most recently, \cite{3DFed} present 3DFed, addressing three prominent defense strategies with corresponding attack modules and introducing an indicator mechanism to assess whether backdoor updates are used in model aggregation. This allows for an adaptive adjustment of the attack strategy. We've also noticed a category of backdoor attacks based on trigger optimization, such as A3FL~\cite{A3FL} and F3BA~\cite{F3BA}. They aim to obtain a robust trigger to make the attacks more covert and persistent. These efforts are compatible with the ones mentioned earlier.

It's important to note that all existing backdoor attacks are data-dependent, meaning they require main task-related data to operate. The development of a data-free backdoor attack remains an open area for exploration.

\subsection{Backdoor Defenses in FL}
\label{sec:backdoor defenses}
Existing defenses against FL backdoors can be categorized into norm constraint-based defenses, outlier detection-based defenses, and consistency detection-based defenses.

Norm constraint-based defenses posit that the optimal point for the backdoor task typically deviates significantly from the optimal point for the main task. This results in the norm of backdoor updates being much larger than that of benign updates. Consequently, these defenses constrain the norm of all local updates within a reasonable range. Norm Clipping~\cite{NormClipping} serves as a representative example of such defenses. Additionally, some other defenses~\cite{FPD,MABRFL,FLTrust} also leverage this characteristic to prevent malicious updates from dominating the global model.

Outlier detection-based defenses assert that backdoor updates and benign updates exhibit substantial differences in their distributions, with benign updates typically being densely distributed. In contrast, backdoor updates can be considered as outliers. Building on this premise, RFLBAT~\cite{RFLBAT} utilizes Principal Component Analysis (PCA) to project local updates into a low-dimensional space. Subsequently, it employs a clustering algorithm to identify outliers, marking them as backdoor updates. FLAME~\cite{FLAME} identifies updates that deviate significantly in direction from the overall trend as backdoor updates and excludes them from the aggregation queue. FLDetector~\cite{FLDetector} exploits the differences between the predicted model and the actual model to discover outliers.

Consistency detection-based defenses argue that all backdoor updates share the same objective, namely, to classify trigger-carrying samples as the target label. Therefore, these updates exhibit strong consistency, either in terms of update directions or neuron activations. On the other hand, diverse benign updates may display lower consistency due to data heterogeneity~\cite{FedProx}. With this understanding, FoolsGold~\cite{Sybils} assigns lower aggregation weights to updates with high pairwise cosine similarities, thereby mitigating the impact of backdoor updates. DeepSight~\cite{DeepSight} uses the consistency on neuron activations in the backdoor model to detect malicious updates.
\section{Preliminaries}
\subsection{Federated Learning}
FL involves iteratively exchanging model parameters between a central server and multiple clients, enabling collaborative training on distributed private data. In broad strokes, FL encompasses the following steps:
\begin{itemize}
	\item \textbf{Step I:} The central server dispatches the global model $w$ to individual clients.
	\item \textbf{Step II:} Each client individually fine-tunes the global model $w$ on their local dataset, resulting in a refined model $w^{\prime}$. Subsequently, the model update $u^{\prime}:=w^{\prime}-w$ is uploaded to the server.
	\item \textbf{Step III:} The server utilizes all the updates (denoted as a set $S$) to enhance the global model as follows:
 \begin{equation}
     w \gets w + AGR(\{u^{\prime}|u^{\prime} \in S\}).
 \end{equation}
 Here, $AGR$ denotes the aggregation algorithm, such as FedAvg~\cite{FedAvg}. 
\end{itemize}

These steps are iteratively performed until a satisfactory global model is obtained. Note that, for the sake of conciseness, we have omitted the iteration round and the client index. 
\section{Threat Model}
\subsection{Adversary's Goal}
The adversary should successfully inject a backdoor without compromising the availability of the global model, even in the presence of SOTA backdoor defenses deployed on the central server. This implies that the adversary needs to concurrently achieve the following three goals:

\begin{itemize}
    \item \textbf{Stealthiness.}
    Backdoor updates must effectively masquerade as benign updates, evading detection by defense schemes to ensure that the backdoor-related knowledge becomes integrated into the global model.
   
    \item \textbf{Fidelity.}
    The backdoor attack should not undermine the global model's ability to recognize clean samples. The post-attack global model must maintain a level of accuracy in the main task comparable to its pre-attack state.
   
    \item \textbf{Effectiveness.} 
     In the case of samples containing the trigger, the global model should exhibit a very high accuracy in recognizing them as the target class, as predetermined by the adversary.
    
\end{itemize}

% Note that the stealthiness goal serves as a prerequisite for both the fidelity and effectiveness goals. Only when the backdoor updates are sufficiently covert to successfully evade the detection by defenses can we meaningfully discuss the efficacy of backdoor injection and the impact of backdoor updates on the overall model accuracy.
\subsection{Adversary's Capability and Knowledge}
We posit that the adversary can emulate a series of fake clients (also referred to as attackers) using open-source projects or Android emulators~\cite{FakeClients} to attain an attacker proportion that aligns with the common academic research scenario, typically around $20\%$. Note that these fake clients lack access to main task-relevant data. The adversary is entirely unaware of the training data and model updates of benign clients, as well as the defense scheme deployed on the server. Furthermore, it cannot disrupt the model training process of benign clients or server decision-making. The sole element within the adversary's control is the training process of the fake clients. The fake clients can mimic the behavior of benign clients to launch covert backdoor attacks.

\section{DarkFed}
\subsection{Motivation Behind DarkFed}
Drawing insights from the results in Tab.~\ref{tab: AcademicVSIndustry}, it becomes apparent that a high attacker proportion is an indispensable prerequisite for the success of backdoor attacks. Nevertheless, achieving such a high attacker proportion is usually unattainable in real-world industrial settings, rendering the existing FL backdoor attack methods impractical. While some literature~\cite{FakeClients} suggests the emulation of a substantial number of fake clients to match the attacker proportion of academic research scenarios, these fake clients are bereft of task-relevant data. Consequently, we are compelled to execute the backdoor attack in a data-free manner.

\subsection{Backdoor Attack with Shadow Dataset}
To realize the data-free backdoor attack, it naturally prompts us to explore the utility of a shadow dataset, because obtaining diverse data unrelated to the main task is quite easy. For example, it can be achieved through web scraping of publicly available data or even generating a series of data points using Gaussian distribution. Consequently, we follow \cite{data-free} and embark on exploring the impact of shadow datasets on backdoor attacks. 
% It's anticipated that the use of a substitute dataset for poisoning will enable the clean model to sustain its initial high ACC while concurrently attaining a notable ASR.

Formally, for a clean model $w$ and a shadow dataset $D_s$, we fine-tune $w$ into a backdoored version $w^\prime$ with the following optimization objective:
\begin{equation}
\begin{aligned}
& \min _{w^{\prime}} L=L_{cl}+\lambda_1 L_{bk}, \\
& L_{cl}=\sum_{x_i \in D_{sc}} \mathcal{L}\left(w^{\prime}\left(x_i\right), w\left(x_i\right)\right), \\
& L_{bk}=\sum_{\widetilde{x}_i \in D_{sp}} \mathcal{L}\left(w^{\prime}\left(\widetilde{x}_i\right), y_t\right),
\end{aligned}
\label{fomula:backdoor}
\end{equation}
where $D_{sc}$ represents the clean dataset without any modifications in the shadow dataset, and $D_{sp}$ represents the poisoned dataset with triggers applied. $\mathcal{L}$ is a loss function, \eg, cross entropy. $w(\cdot)$ and $w^{\prime}(\cdot)$ denote the logits of $w$ and $w^{\prime}$, respectively, $y_t$ is the target label, and the hyperparameter $\lambda_1$ is used to balance the model performance and the poisoning effect. The purpose of $L_{cl}$ is to maintain the performance of the main task, while $L_{bk}$ aims to learn the knowledge related to the backdoor. It should be noted that throughout the entire fine-tuning process, $\lambda_1$ is fixed at 1, and the clean model $w$ is treated as a constant that remains unchanged.

We consider three popular image classification tasks: CIFAR-10~\cite{CIFAR10}, CIFAR-100~\cite{CIFAR10}, and GTSRB~\cite{GTSRB}. As for the shadow datasets, in addition to these three real datasets, we also include three synthetic datasets constructed based on certain distributions. These synthetic datasets do not contain any semantic information. Due to the commonality of the three real datasets, we omit their introduction here and focus solely on describing the three synthetic datasets.

\begin{itemize}
    \item \textbf{Gauss-I.}
    % Gauss-I comprises $20k$ images generated according to a Gaussian distribution, each with dimensions of $32\times32\times3$. For each image, we employ the Gaussian distribution $N(0.5, 1^2)$ to generate pixel values for each dimension. Any values exceeding 1 or falling below 0 are appropriately clipped to ensure adherence to the valid range.
    Each image is of size $32\times32\times3$, with each pixel value generated from a Gaussian distribution $N(0.5, 1^2)$ and located within the range $\left[0,1\right]$.
   
    \item \textbf{Gauss-II.}
    % Gauss-II also comprises $20k$ images generated according to a Gaussian distribution, each with dimensions of $32\times32\times3$. The only distinction between Gauss-II and Gauss-I lies in its utilization of a smaller standard deviation of 0.2.
    Each image is of size $32\times32\times3$, with each pixel value generated from a Gaussian distribution $N(0.5, 0.2^2)$ and located within the range $\left[0,1\right]$.
   
    \item \textbf{Uniform.} 
     % Uniform consists of $20k$ images generated according to a uniform distribution, each with dimensions of $32\times32\times3$. For each image, we employ the uniform distribution $U(0,1)$ to generate pixel values for each dimension.
     Each image is of size $32\times32\times3$, with each pixel value generated from a uniform distribution $U(0, 1)$.
\end{itemize}

\begin{figure}[t]
	\centering
	\subfigure[CIFAR-10]{\label{fig:lf_m}
		\includegraphics[width=0.3\columnwidth]{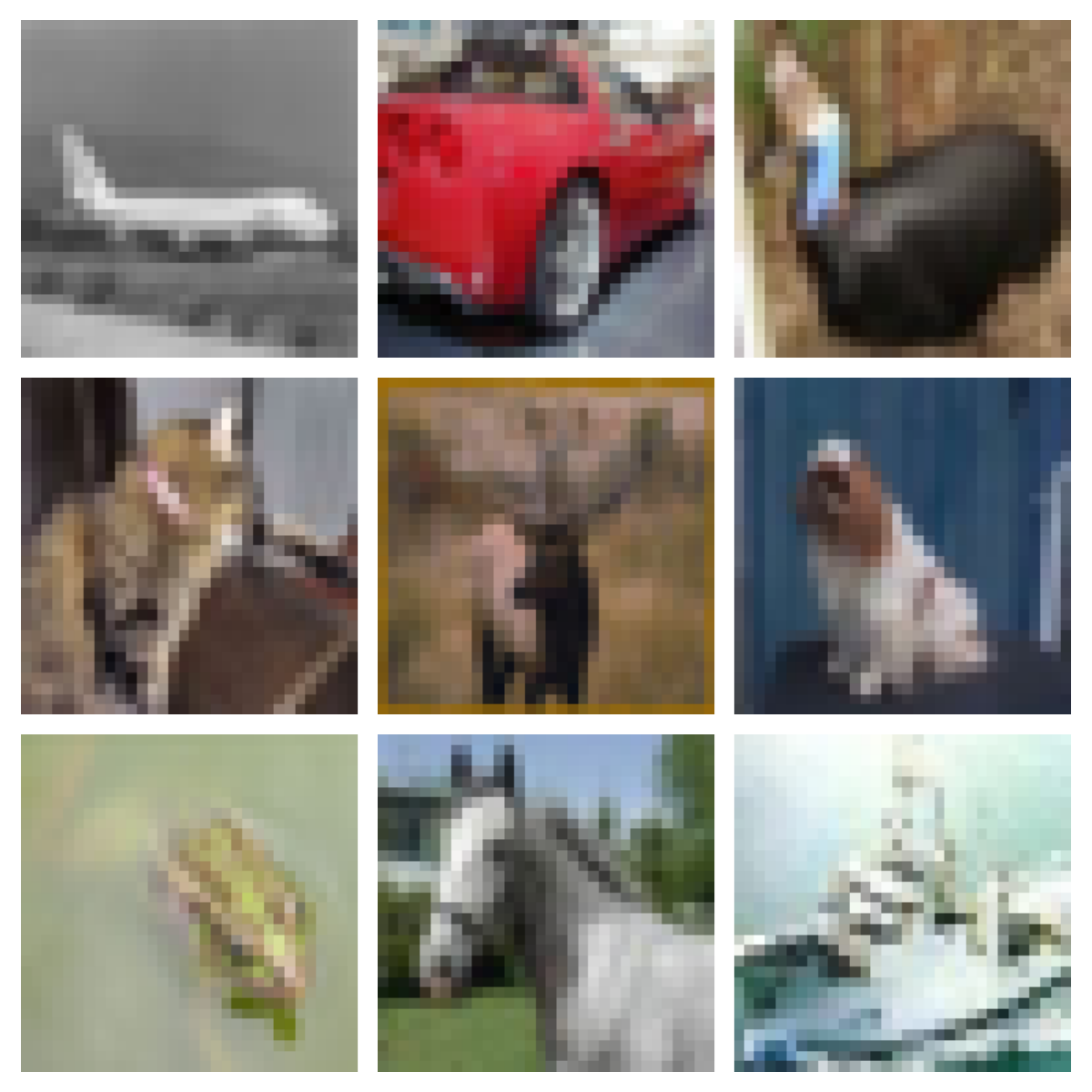}}
	\subfigure[CIFAR-100]{\label{fig:lf_c}
		\includegraphics[width=0.3\columnwidth]{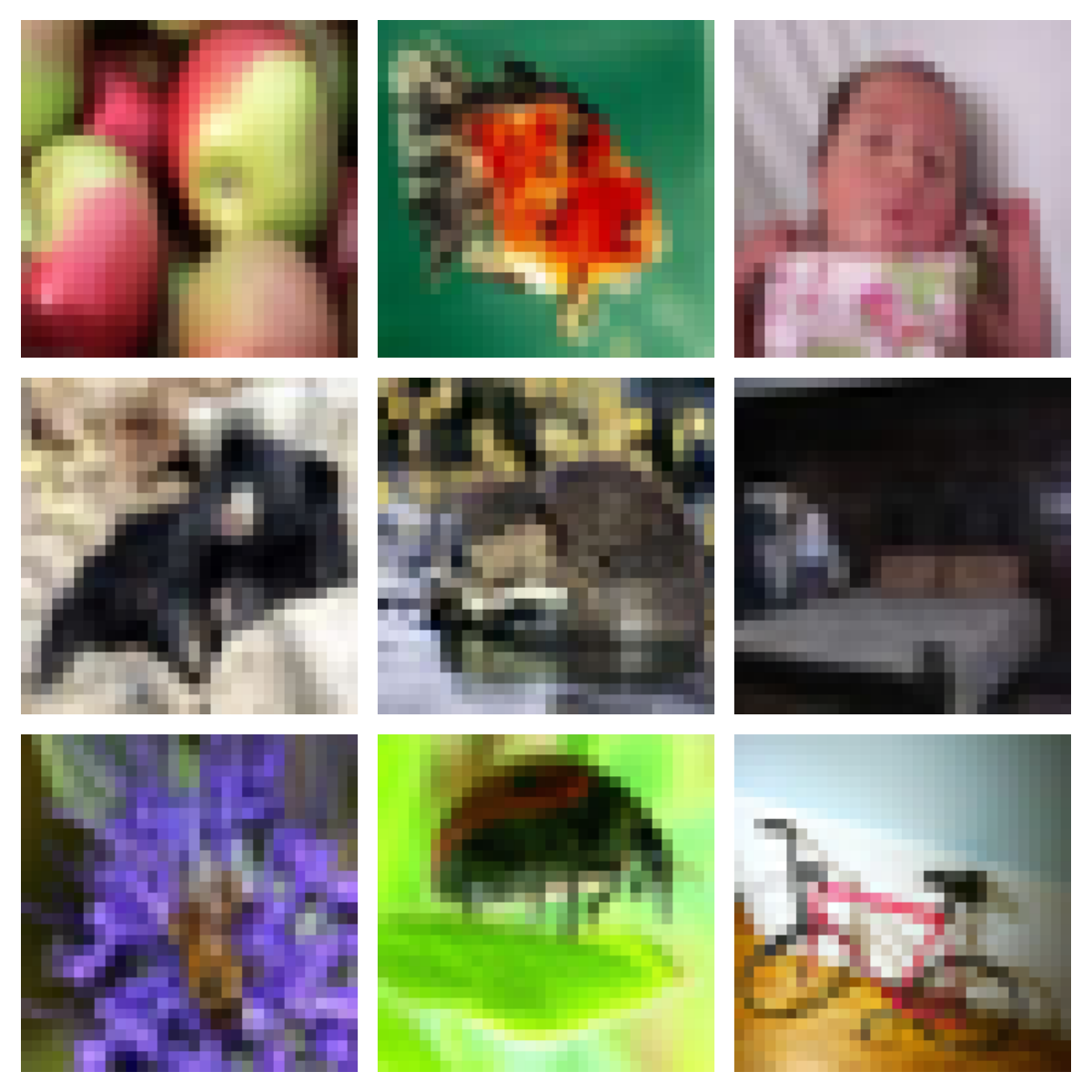}}
  \subfigure[GTSRB]{\label{fig:lf_c}
		\includegraphics[width=0.3\columnwidth]{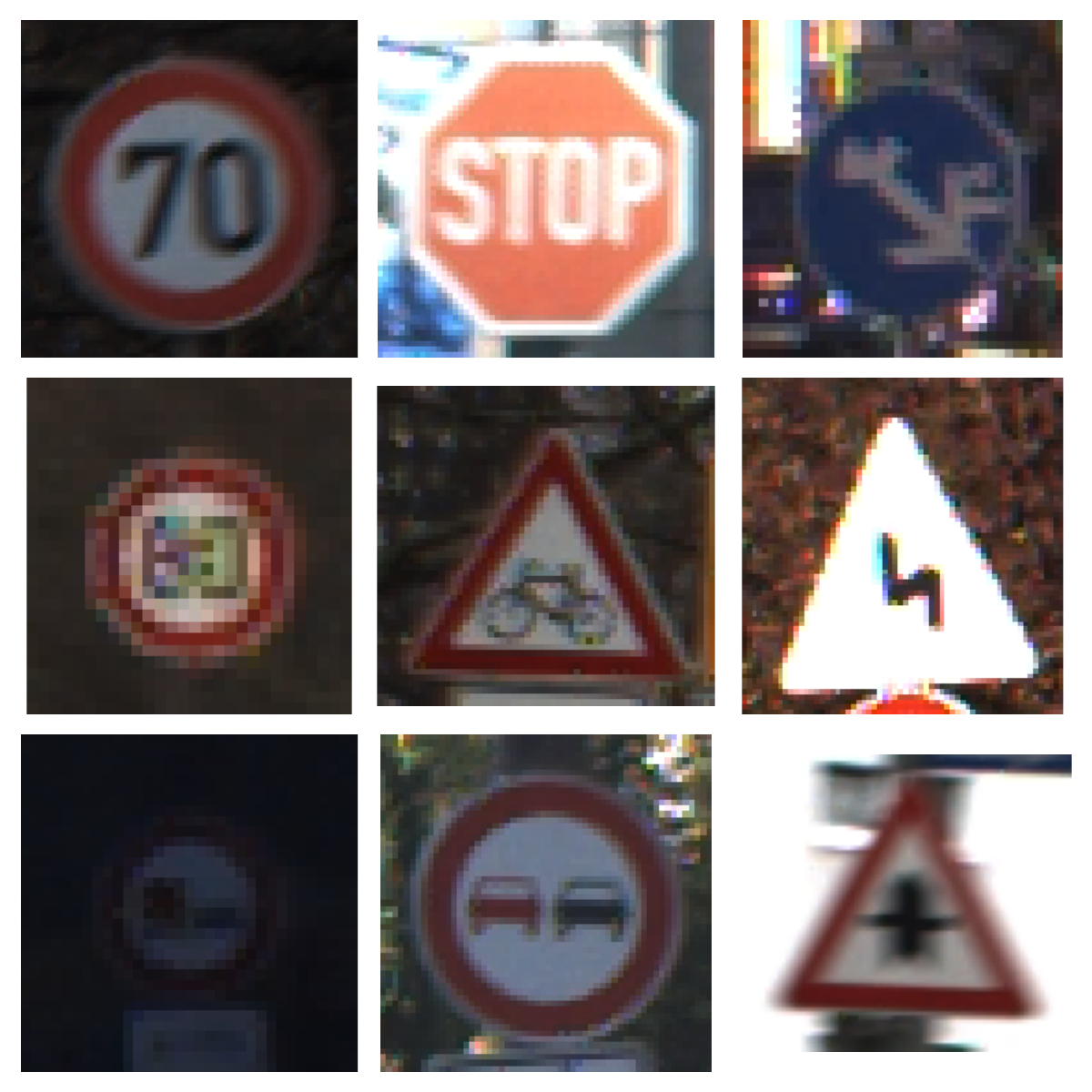}}
  \subfigure[Gauss-I]{\label{fig:lf_c}
		\includegraphics[width=0.3\columnwidth]{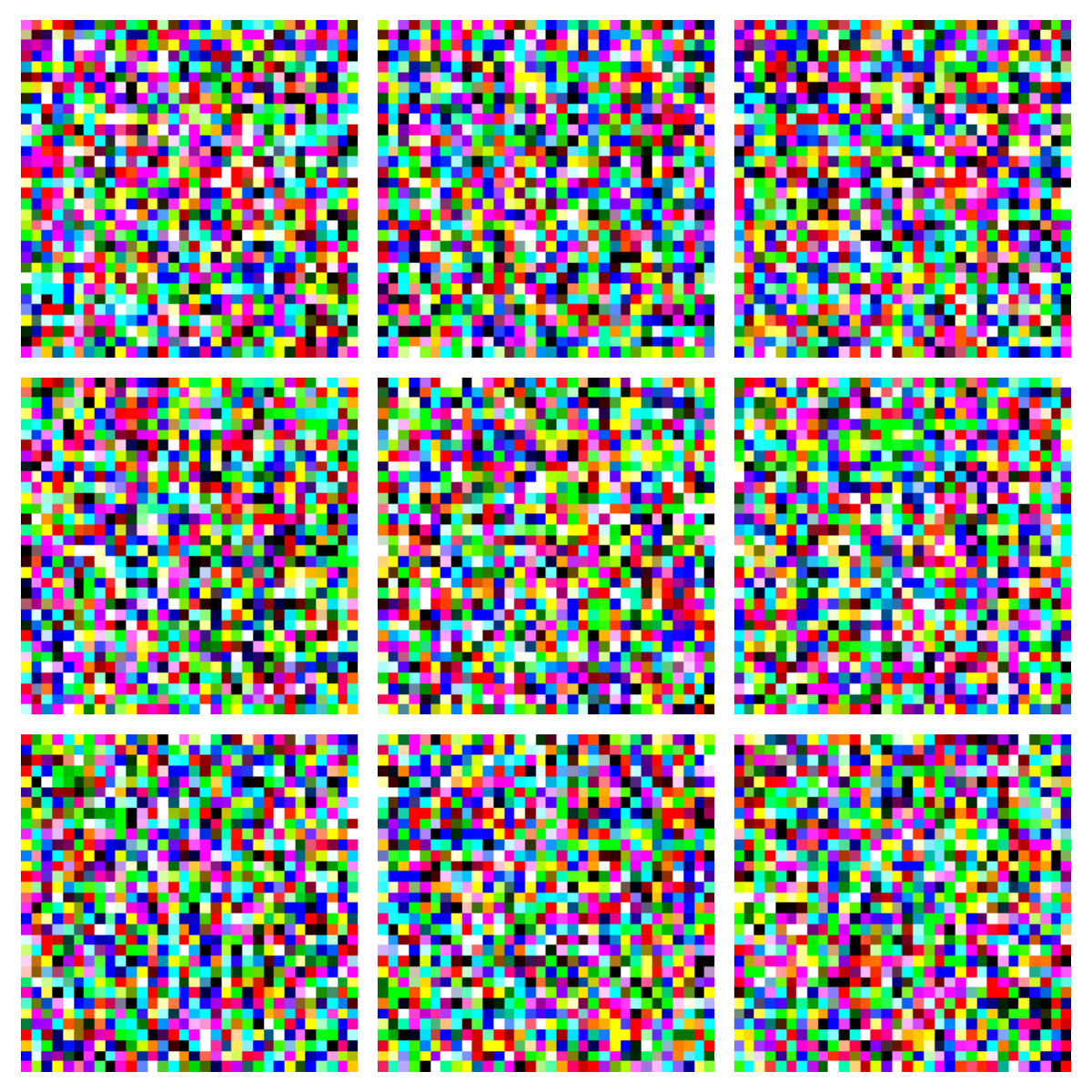}}
  \subfigure[Gauss-II]{\label{fig:lf_c}
		\includegraphics[width=0.3\columnwidth]{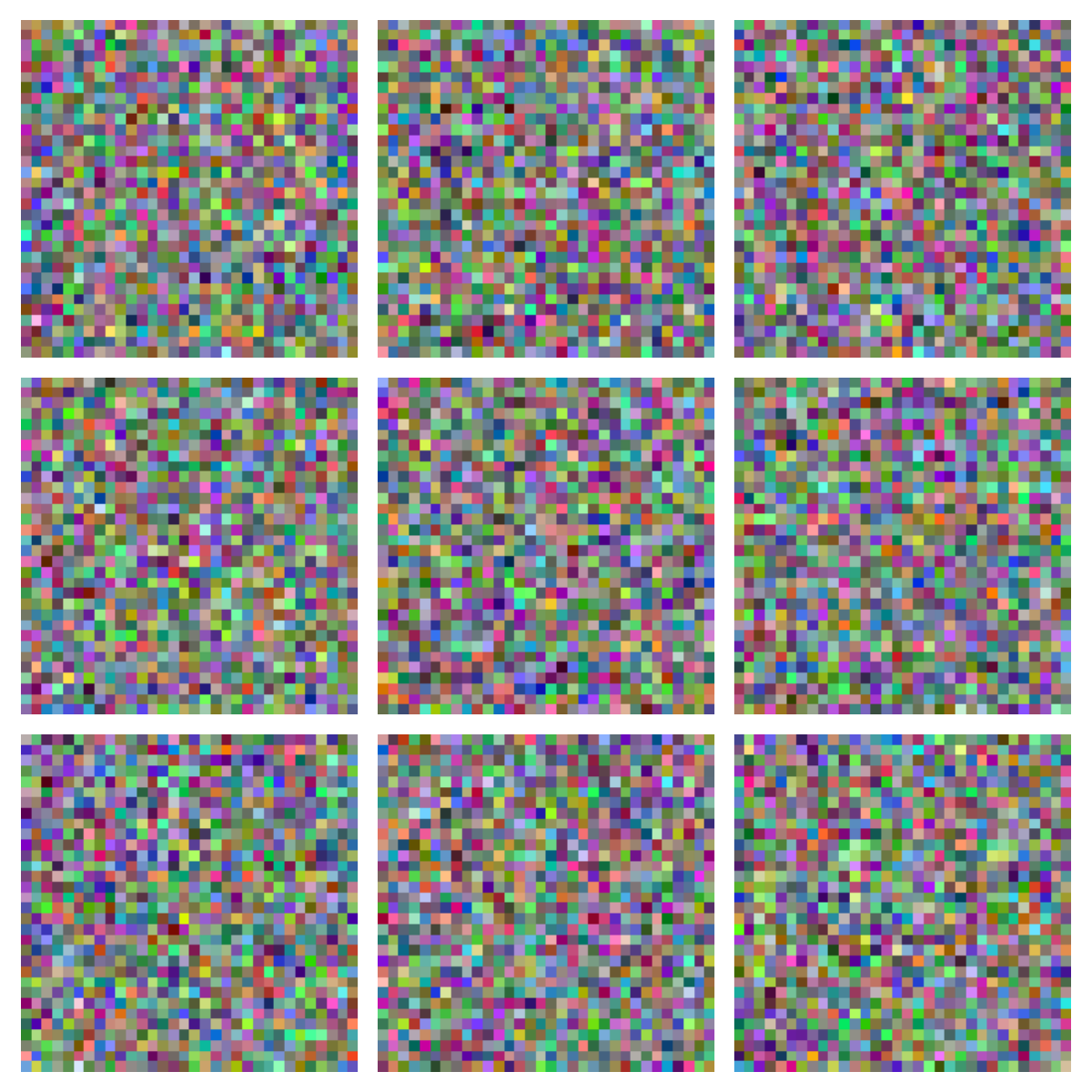}}
  \subfigure[Uniform]{\label{fig:lf_c}
		\includegraphics[width=0.3\columnwidth]{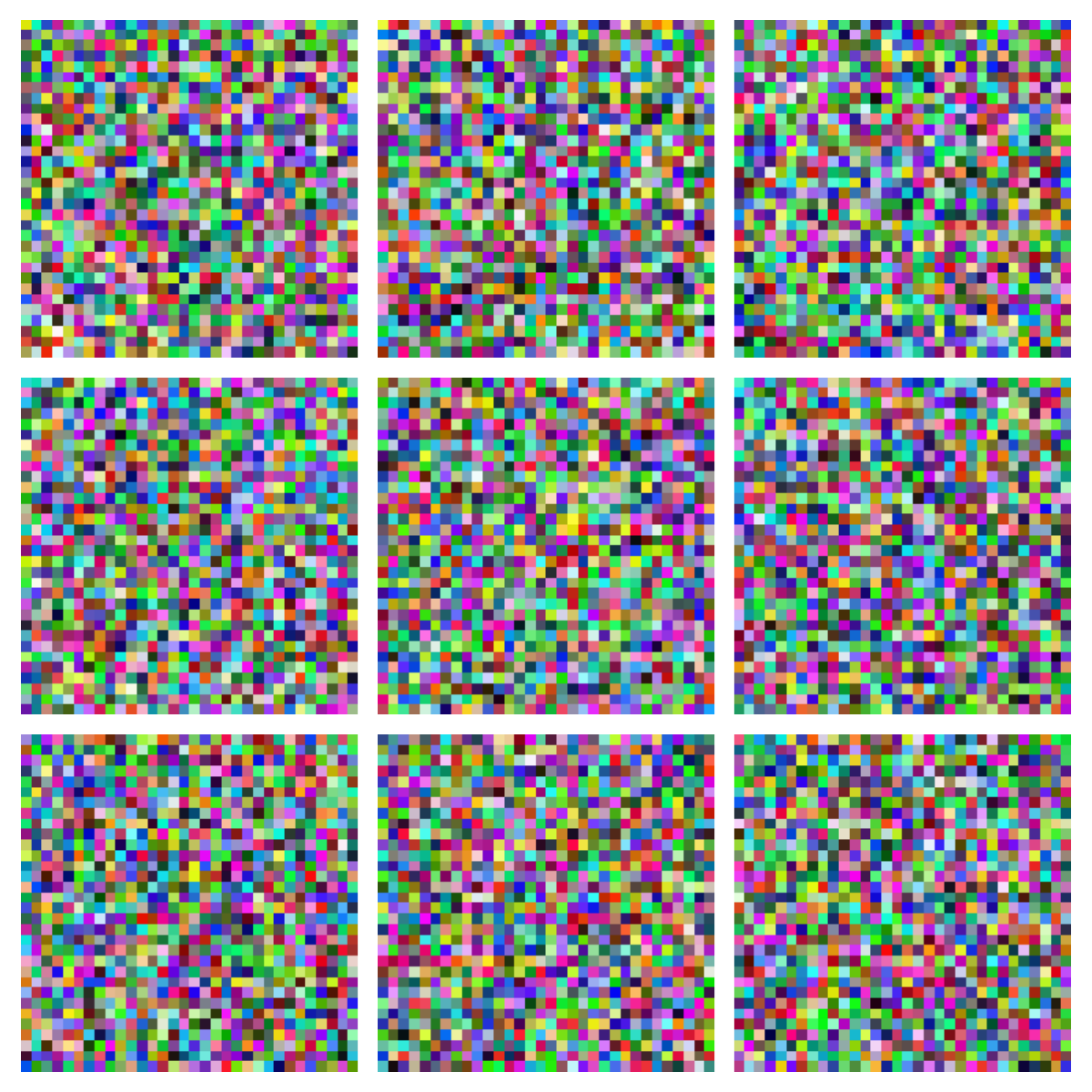}}
    \vspace{-1.5mm}
	\caption{Visual comparison of the shadow datasets.}
   \vspace{-1.5mm}
	\label{fig:visual comparison}
\end{figure}

Fig.~\ref{fig:visual comparison} showcases a subset of samples from these datasets. Notably, significant visual disparities exist among them,
% , except for the similarity between CIFAR-10 and CIFAR-100. GTSRB primarily comprises well-structured traffic signs, while CIFAR-10/100 presents a more diverse array of data, including animals and various modes of transportation. Conversely, the samples in Gauss-I, Gauss-II, and Uniform appear chaotic, resembling random noise. 
which potentially leads to the presumption that their utilization as shadow datasets would lead to a reduction in the model's main task recognition accuracy (\ie, ACC) or an unsatisfactory level of backdoor task accuracy (\ie, ASR). However, to our surprise, as shown in Tab.~\ref{tab: SubstituteDataset}, the shadow dataset, compared to directly using task-related datasets (highlighted in green), does not significantly impact the backdoor performance, especially when utilizing real datasets. The last three rows in the table illustrate the scenario when synthetic datasets serve as shadow datasets. It can be observed that the performance of Gauss-I is comparable to that of real datasets. Gauss-II fails to achieve satisfactory ASR, and we speculate this is due to its small standard deviation during construction, resulting in lower data richness and thus poorer performance. Uniform's performance falls between Gauss-I and Gauss-II, achieving over $90\%$ ASR on each dataset.

We posit that the ability to maintain high ACC with a shadow dataset lies in our strategy of ensuring similarity between the logits of $w^{\prime}$ and $w$ (see $L_{cl}$ in Eq.~(\ref{fomula:backdoor})), rather than employing hard labels. This allows for minor adjustments to be made to $w^{\prime}$ based on $w$. The achievement of high ASR is attributed to the fact that backdoor learning focuses on the mapping relationship between the trigger and the target label, making it less influenced by the shadow dataset itself.

\begin{table}[!t]
\renewcommand\arraystretch{1}
\centering
% \vspace{-4mm}
\caption{Impact of shadow datasets on backdoor performance.}
  \vspace{-1.5mm}
\resizebox{\linewidth}{!}{
\begin{tabular}{|c||cc|cc|cc|}
\hline
 & \multicolumn{2}{c|}{\textbf{CIFAR-10}} & \multicolumn{2}{c|}{\textbf{CIFAR-100}} & \multicolumn{2}{c|}{\textbf{GTSRB}} \\ \cline{2-7} 
\multirow{-2}{*}{\textbf{\begin{tabular}[c]{@{}c@{}}Shadow \\ Dataset\end{tabular}}} & \multicolumn{1}{c|}{\textbf{ACC (\%)}} & \textbf{ASR (\%)} & \multicolumn{1}{c|}{\textbf{ACC (\%)}} & \textbf{ASR (\%)} & \multicolumn{1}{c|}{\textbf{ACC (\%)}} & \textbf{ASR (\%)} \\ \hline
CIFAR-10 & \multicolumn{1}{c|}{\cellcolor[HTML]{9AFF99}89.17} & \cellcolor[HTML]{9AFF99}100.00 & \multicolumn{1}{c|}{78.97} & 100.00 & \multicolumn{1}{c|}{93.17} & 100.00 \\ \hline
CIFAR-100 & \multicolumn{1}{c|}{89.14} & 100.00 & \multicolumn{1}{c|}{\cellcolor[HTML]{9AFF99}79.09} & \cellcolor[HTML]{9AFF99}100.00 & \multicolumn{1}{c|}{93.13} & 100.00 \\ \hline
GTSRB & \multicolumn{1}{c|}{88.93} & 99.81 & \multicolumn{1}{c|}{78.46} & 100.00 & \multicolumn{1}{c|}{\cellcolor[HTML]{9AFF99}93.34} & \cellcolor[HTML]{9AFF99}100.00 \\ \hline
Gauss-I & \multicolumn{1}{c|}{88.90} & 98.36 & \multicolumn{1}{c|}{78.06} & 96.72 & \multicolumn{1}{c|}{93.08} & 99.35 \\ \hline
Gauss-II & \multicolumn{1}{c|}{89.02} & 79.63 & \multicolumn{1}{c|}{78.19} & 75.25 & \multicolumn{1}{c|}{92.60} & 83.54 \\ \hline
Uniform & \multicolumn{1}{c|}{89.06} & 93.19 & \multicolumn{1}{c|}{78.17} & 95.45 & \multicolumn{1}{c|}{93.23} & 97.99 \\ \hline
\end{tabular}
}
% \vspace{-5mm}
\label{tab: SubstituteDataset}
\end{table}

\begin{figure}[t]
	\centering
	\includegraphics[width=1\columnwidth]{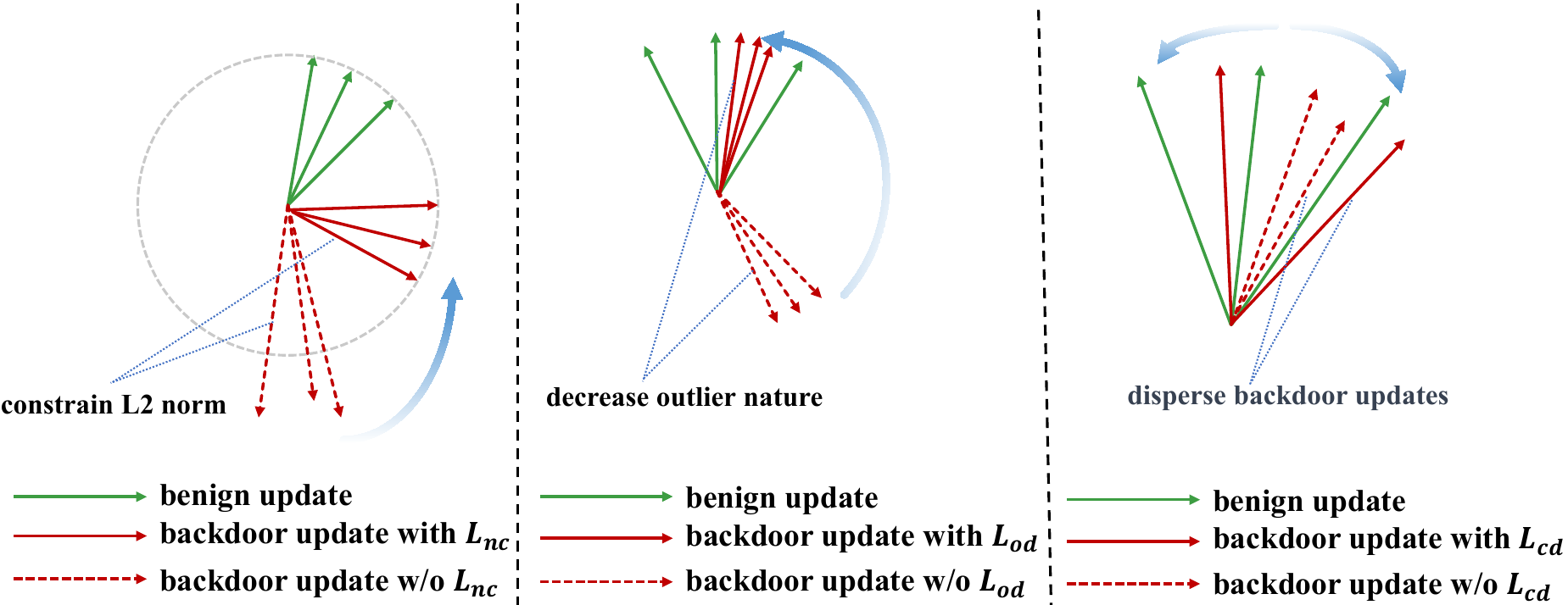}
   \vspace{-1.5mm}
	\caption{Illustration of property mimicry.}
  \vspace{-3mm}
	\label{fig: illustation}
\end{figure}

\subsection{Property Mimicry}
The preceding exploration indicates that leveraging a shadow dataset and Eq.~(\ref{fomula:backdoor}) can simultaneously achieve high ACC (the fidelity goal) and high ASR (the effectiveness goal). Therefore, a naive idea is to directly execute this on fake clients. However,
% the FL system can deploy corresponding defenses to prevent the global model from being maliciously injected with a backdoor.
this approach is easily thwarted by existing defense mechanisms, leading to a failure in backdoor implantation (\ie, the stealthiness goal is not achieved). To enhance stealthiness, our core idea is to make the backdoor updates generated by fake clients mimic benign updates in terms of properties (\eg, magnitude, distribution, and consistency). This approach makes it challenging to distinguish between these two types of updates based on certain properties, thereby evading defense mechanisms.

\noindent\textbf{Moderate magnitude.} The magnitude of benign updates is typically moderate, and this feature is also exploited by norm constraint-based defenses to filter or constrain backdoor updates. To mimic this property of benign updates, an intuitive approach is to prevent the magnitude of backdoor updates from becoming excessively large, and this can be effortlessly achieved by incorporating a constraint term. 

Formally, for the global model $w$ and the local model $w^{\prime}$, we consider the following constraint term:
\begin{equation}
    L_{nc} = ||w^{\prime}-w||_2.
    \label{eq:lnc}
\end{equation}
The inspiration for this constraint term is drawn from FedProx~\cite{FedProx}, which also incorporates an identical constraint term. However, our approach differs fundamentally from FedProx. Firstly, FedProx employs this constraint term to alleviate the issue of model accuracy degradation caused by statistical heterogeneity in FL, whereas our objective is to enhance the stealthiness of backdoor attacks. Furthermore, FedProx applies this constraint term to all clients, whereas we restrict its use solely to malicious clients. While this constraint term is simple and intuitive, it effectively restricts the magnitude of backdoor updates without compromising the efficiency of backdoor injection or the backdoor accuracy. We surmise that $L_{nc}$ can guide the backdoor model to search for a joint optimal point near the global model, where both the backdoor task and the primary task perform well. The effect of $L_{nc}$ is illustrated in Fig.~\ref{fig: illustation} (left).

\noindent\textbf{Reasonable distribution. }
Benign updates typically exhibit a reasonable distribution, so when malicious updates are introduced, they can be detected as outliers. To mimic this characteristic of benign updates, a favorable countermeasure involves narrowing the similarity between backdoor updates and benign updates. This confuses the defense mechanism, making it challenging to detect \textit{out-of-distribution} (OOD) values and potentially leading to erroneous identifications (\ie, classifying benign updates as malicious). Considering the well-established capability of cosine similarity in measuring update similarity, and its widespread adoption in various defense methods~\cite{FLAME,FLTrust}, we employ it as a metric to quantify the similarity between malicious and benign updates. Nevertheless, the adversary lacks knowledge of benign models.  Although it can train a set of emulated benign models using the clean shadow dataset $D_{sc}$ and $L_{cl}$ with Eq.~(\ref{fomula:backdoor}), the emulated benign models may deviate substantially from real benign ones. This discrepancy increases the risk of backdoor models veering in the wrong direction, thereby amplifying the visibility of the attack.

In this context, we propose employing the \textit{double exponential smoothing} (DES) algorithm to predict the forthcoming round's global model because DES has exhibited remarkable effectiveness in predicting the distribution of benign models~\cite{RobustFL}.
Formally, for the global model $w$ and the local model $w^{\prime}$, we introduce the following constraint term:
\begin{equation}
    L_{od} = (\cos(w^{\prime}-w,\hat{w}-w)-\alpha)^2,
    \label{eq:lod}
\end{equation}
where $\hat{w}$ denotes the predicted global model through DES, $\cos$ denotes cosine similarity, and the hyperparameter $\alpha$ represents the estimated cosine similarity between benign updates. The constraint term $L_{od}$ tightens the similarity between backdoor updates and benign updates, preventing the backdoor updates from resembling OOD values. This allows them to circumvent outlier detection-based defenses. The effect of $L_{od}$ is illustrated in Fig.~\ref{fig: illustation} (middle).

% Formally, for the global model $w$ and the local model $w^{\prime}$, we consider the following constraint term:
% \begin{equation}
%     L_{od} = \sum_{w^{\prime\prime}\in W_{eb}}||cos(w^{\prime}-w,w^{\prime\prime}-w)-\alpha||
% \end{equation}
% where $W_{eb}$ is a collection of emulated benign models trained on the clean substitute dataset $D_{sc}$ using $L_0$ in Eq.~(\ref{fomula:backdoor}), these models are employed to emulate the distribution of real benign models. $cos$ denotes cosine similarity, and $\alpha$ represents the average cosine similarity between all pairs of models within $W_{eb}$. The constraint term $L_{od}$ tightens the similarity between backdoor updates and benign updates, preventing the backdoor updates from resembling OOD values. This allows them to circumvent outlier detection-based defenses.

% \begin{figure}[t]
% 	\centering
%     \subfigure[$L_{nc}$]{\label{fig:lnc}
% 		\includegraphics[width=0.32\columnwidth]{figs/evading nc.pdf}}
% 	\subfigure[$L_{od}$]{\label{fig:lod}
% 		\includegraphics[width=0.3\columnwidth]{figs/evading od.pdf}}
% 	\subfigure[$L_{cd}$]{\label{fig:lcd}
% 		\includegraphics[width=0.3\columnwidth]{figs/evading cd.pdf}}
% 	\caption{Illustration of our designed constraint terms}
%  % \vspace{-5mm}
% 	\label{fig: illustation}
% \end{figure}

\begin{algorithm}[t]
\SetKwInOut{KIN}{Input}
\SetKwInOut{KOUT}{Output}
\caption{A Complete Description of DarkFed}
\label{Alg:DarkFed}
\KIN{$D_{s}$: shadow dataset; $w$: global model; $W_{bk}$: collection of all backdoor models; $\alpha$: estimated cosine similarity between benign updates; $E$: local epoch; $B$: batch size; $\lambda$: coefficient balancing stealthiness with fidelity and effectiveness; $\eta$: learning rate.}
\KOUT{the backdoored models $W_{bk}$}
// Initialize all backdoor models with global model\\
\For{$w^{\prime}\in W_{bk}$}{$w^{\prime} \gets w$}
Obtain the predicted global model $\hat{w}$ through DES\\
$\mathcal{B} \gets$ (split $D_{s}$ into batches of size $B$)\\
// Optimize each backdoor model following Eq.~(\ref{eq:final objective})\\
\For{each epoch $e \in [1, E]$}{
\For{$w^{\prime} \in W_{bk}$}{
\For{each batch $b \in \mathcal{B}$}{
% // Calculate loss value $L$ following Eq.~(\ref{eq:final objective})\\
 $L=L_{cl}+ L_{bk}+ \lambda (L_{nc} + L_{od} + L_{cd})$\\
$w^{\prime} \gets w^{\prime}-\eta \nabla_{w^{\prime}}L$
}
}
}
\end{algorithm}

\noindent\textbf{Limited consistency. }The consistency of benign updates is typically limited, as they do not share the same objective, unlike backdoor updates. This principle forms the core perspective of consistency detection-based defenses. To simulate this property of benign updates, we aim to optimize the representation of backdoor updates, ensuring they demonstrate a consistency akin to benign updates. Specifically, for the global model $w$ and the local model $w^{\prime}$, we introduce the following constraint term:
\begin{equation}
    L_{cd} = \sum_{w^{\prime\prime}\in W_{bk}-\{w^{\prime}\}}(\cos(w^{\prime}-w,w^{\prime\prime}-w)-\alpha)^2,
    \label{eq:lcd}
\end{equation}
where $W_{bk}$ represents the collection of all backdoor models, and $\alpha$ carries the same meaning as in Eq.~(\ref{eq:lod}). The constraint term $L_{cd}$ reduces the consistency of backdoor updates to a level comparable to benign updates, effectively confounding defensive strategies. The effect of $L_{cd}$ is illustrated in Fig.~\ref{fig: illustation} (right).

\subsection{A Complete Description of DarkFed}
Combining constraint terms (\ref{eq:lnc}), (\ref{eq:lod}), and (\ref{eq:lcd}), we can reformulate the optimization objective in (\ref{fomula:backdoor}) as follow:
\begin{equation}
    \min _{w^{\prime}} L=L_{cl}+ L_{bk}+ \lambda (L_{nc} + L_{od} + L_{cd}),
    \label{eq:final objective}
\end{equation}
% \begin{equation}
%     \min _{w^{\prime}} L=L_{cl}+\lambda_1 L_{bk}+ \lambda_2 L_{nc} +\lambda_3  L_{od} +\lambda_4 L_{cd}
%     \label{eq:final objective}
% \end{equation}
where $L_{cl}$ is designed to achieve the fidelity goal. $L_{bk}$ aims to fulfill the effectiveness goal. $L_{nc}$, $L_{od}$, and $L_{cd}$ contribute to realizing the stealthiness goal. $\lambda$ is a coefficient balancing stealthiness with fidelity and effectiveness. Note that this optimization objective is extensible. While it currently encompasses evasion constraint terms designed for existing defense categories, new defenses may emerge in the future that do not fall into any of these categories. In such cases, we can still design corresponding constraint terms to extend this optimization objective.

Alg.~\ref{Alg:DarkFed} provides a complete description of the DarkFed scheme during a round of global iteration. Upon receiving the global model from the central server, all fake clients initialize their local models with this global model (Lines 1-3). Subsequently, the DES algorithm is employed to obtain the predicted global model (Line 4), which is utilized in the computation of $L_{od}$ (refer to Eq.~(\ref{eq:lod})). Then the shadow dataset is divided into multiple batches for local training (Line 5). Finally, optimization is performed for each backdoor model based on Eq.~(\ref{eq:final objective}) (Lines 6-11).

\section{Experiments}
\subsection{Experimental Setup}
\textbf{Datasets, models, and codes.}
We consider three multi-channel image classification datasets: CIFAR-10~\cite{CIFAR10}, CIFAR-100~\cite{CIFAR10}, and GTSRB~\cite{GTSRB}. Because, compared to single-channel datasets like MNIST~\cite{MNIST} and Fashion-MNIST~\cite{Fashion-MNIST}, multi-channel datasets are more complex and better represent real-world scenarios. For CIFAR-10 and CIFAR-100, we employ ResNet-18 as the model structure. For GTSRB, we construct a VGG-like model as the global model. It's worth noting that, to expedite experimentation, we follow ~\cite{3DFed}, employing a pre-trained model as the initial global model to simulate a scenario where the global model is nearing convergence. The initial model's ACC and ASR on the three datasets are provided in Tab.~\ref{tab: Initial model performance}. Note that when calculating ASR, samples corresponding to the target label have not been excluded. This results in some backdoor samples being identified as the target class not because they are triggered, but because they inherently belong to the target class. Consequently, this leads to a relatively higher ASR. This approach is justified since backdooring a FL system when the global model is close to convergence is enough. Our codes will be available at \href{https://github.com/hustweiwan/DarkFed}{https://github.com/hustweiwan/DarkFed}.

\noindent\textbf{Shadow datasets.}
% As illustrated in Fig~\ref{fig:Substitute datasets}, 
For the CIFAR-10 and CIFAR-100 classification tasks, we employ GTSRB as the shadow dataset. Conversely, for the GTSRB classification task, we utilize CIFAR-100 as the shadow dataset. This decision is guided by the relatively small domain gap between CIFAR-10 and CIFAR-100, and we avoid using them interchangeably as shadow datasets.

\begin{table}[!t]
\renewcommand\arraystretch{1}
\centering
% \vspace{-4mm}
\caption{Performance of the initial global model.}
  \vspace{-1.5mm}
\resizebox{\linewidth}{!}{
\begin{tabular}{|cc|cc|cc|}
\hline
\multicolumn{2}{|c|}{\textbf{CIFAR-10}} & \multicolumn{2}{c|}{\textbf{CIFAR-100}} & \multicolumn{2}{c|}{\textbf{GTSRB}} \\ \hline
\multicolumn{1}{|c|}{\textbf{ACC (\%)}} & \textbf{ASR (\%)} & \multicolumn{1}{c|}{\textbf{ACC(\%)}} & \textbf{ASR(\%)} & \multicolumn{1}{c|}{\textbf{ACC (\%)}} & \textbf{ASR (\%)} \\ \hline
\multicolumn{1}{|c|}{90.15} & 8.77 & \multicolumn{1}{c|}{79.01} & 0.79 & \multicolumn{1}{c|}{93.08} & 2.93 \\ \hline
\end{tabular}
}
\vspace{-2mm}
\label{tab: Initial model performance}
\end{table}

\begin{table}[!t]
\renewcommand\arraystretch{1}
\centering
% \vspace{-4mm}
\caption{Patameter settings in Alg.~\ref{Alg:DarkFed}.}
  \vspace{-1.5mm}
\resizebox{\linewidth}{!}{
\begin{tabular}{|c||c|c|c|c|c|c|}
\hline
\textbf{Dataset} & \textbf{$\alpha$} & \textbf{$\lambda$} & \textbf{$\eta$} & \textbf{$E$} & \textbf{$B$} & \textbf{$D_s$} \\ \hline
CIFAR-10 & 0 & 0.5 & 0.005 & 15 & 64 & GTSRB \\ \hline
CIFAR-100 & 0 & 0.5 & 0.001 & 15 & 64 & GTSRB \\ \hline
GTSRB & 0 & 0.5 & 0.00005 & 15 & 64 & CIFAR-100 \\ \hline
\end{tabular}
}
\vspace{-5mm}
\label{tab: Patameter settings}
\end{table}
% \begin{figure}[t]
% \centerline{\includegraphics[width=1\columnwidth]{figs/substitute dataset.pdf}}
% \caption{Substitute datasets}
% \label{fig:Substitute datasets}
% \end{figure}

\noindent\textbf{Attack settings.}
In line with \cite{CerP}, we establish a FL system encompassing 100 clients, with $20\%$ of them being emulated fake clients. These fake clients lack training data relevant to the main task and instead utilize a publicly scraped dataset or a synthetic dataset (\eg, generated through Gaussian distribution) to introduce a backdoor. $20\%$ of the total clients are randomly selected in each iteration. The parameter settings in Alg.~\ref{Alg:DarkFed} are delineated in Tab~\ref{tab: Patameter settings}. 
% Notably, the parameters for each dataset are nearly identical, with the exception of the learning rate $\eta$. This variation stems from GTSRB being relatively easier to learn than CIFAR-10/100, necessitating a smaller learning rate. The consistency in parameter settings across datasets underscores DarkFed's insensitivity to both the dataset and model, enhancing its deployability. 
One might wonder why the estimated cosine similarity between benign updates (\ie, $\alpha$) consistently remains at 0 for different datasets. This is attributed to the research in \cite{MABRFL}, which indicates that benign updates exhibit similarity only in the initial rounds, becoming nearly orthogonal in subsequent iterations.

\begin{figure*}[t]
	\centering
	\subfigure[FedAvg]{
		\includegraphics[width=0.38\columnwidth]{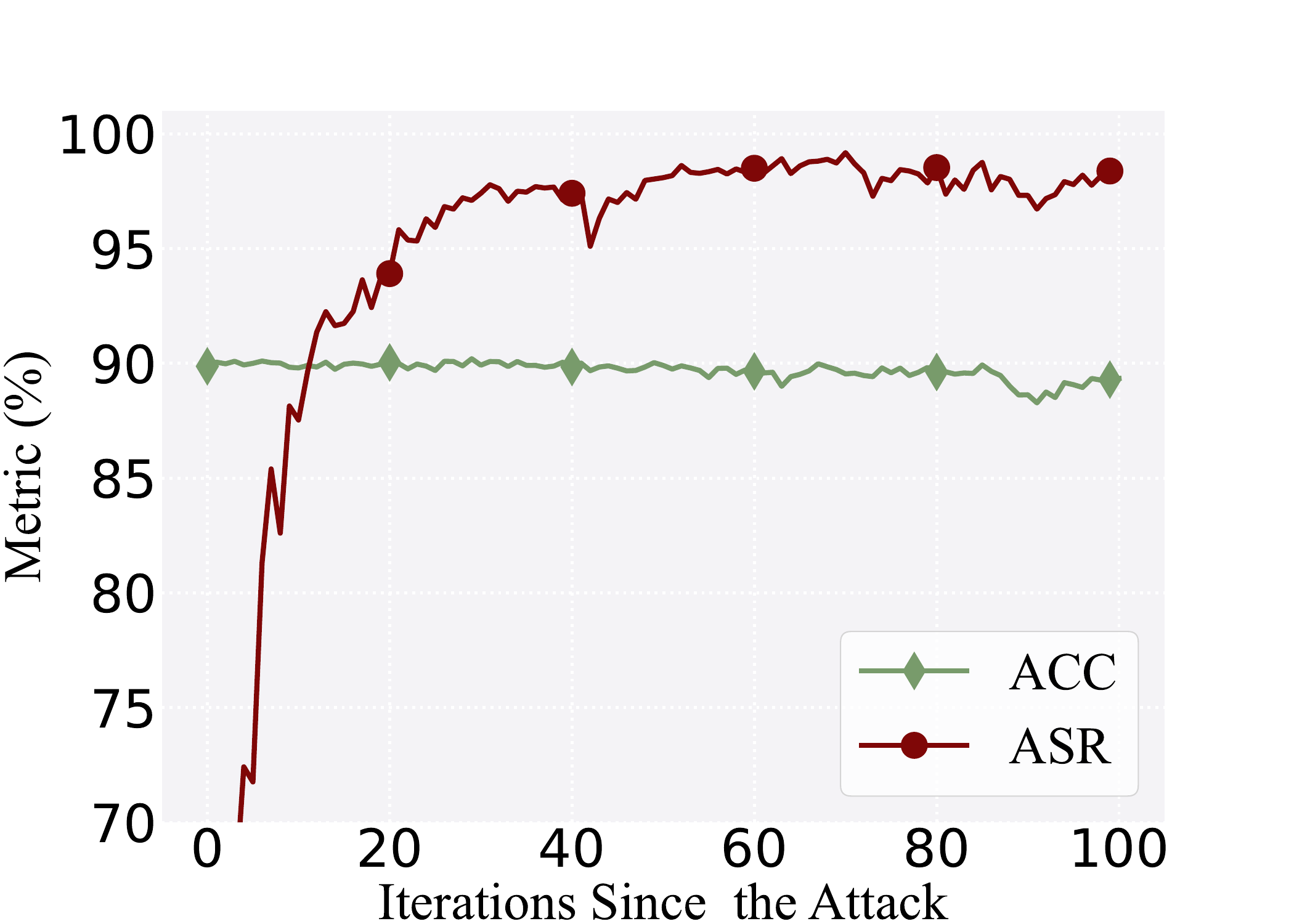}}
	\subfigure[Norm Clipping]{\label{fig:lf_c}
		\includegraphics[width=0.38\columnwidth]{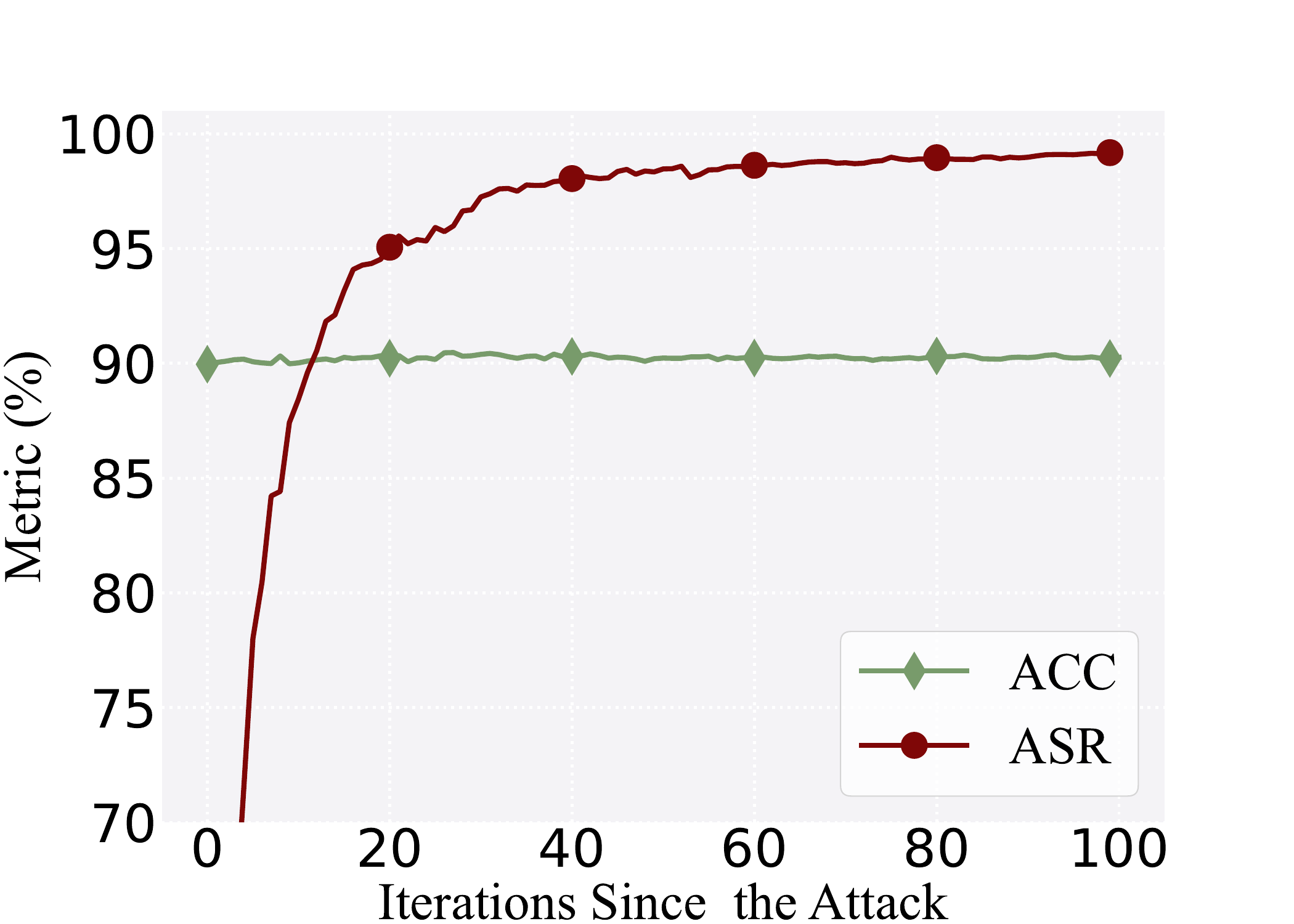}}
  \subfigure[FLAME]{\label{fig:lf_c}
		\includegraphics[width=0.38\columnwidth]{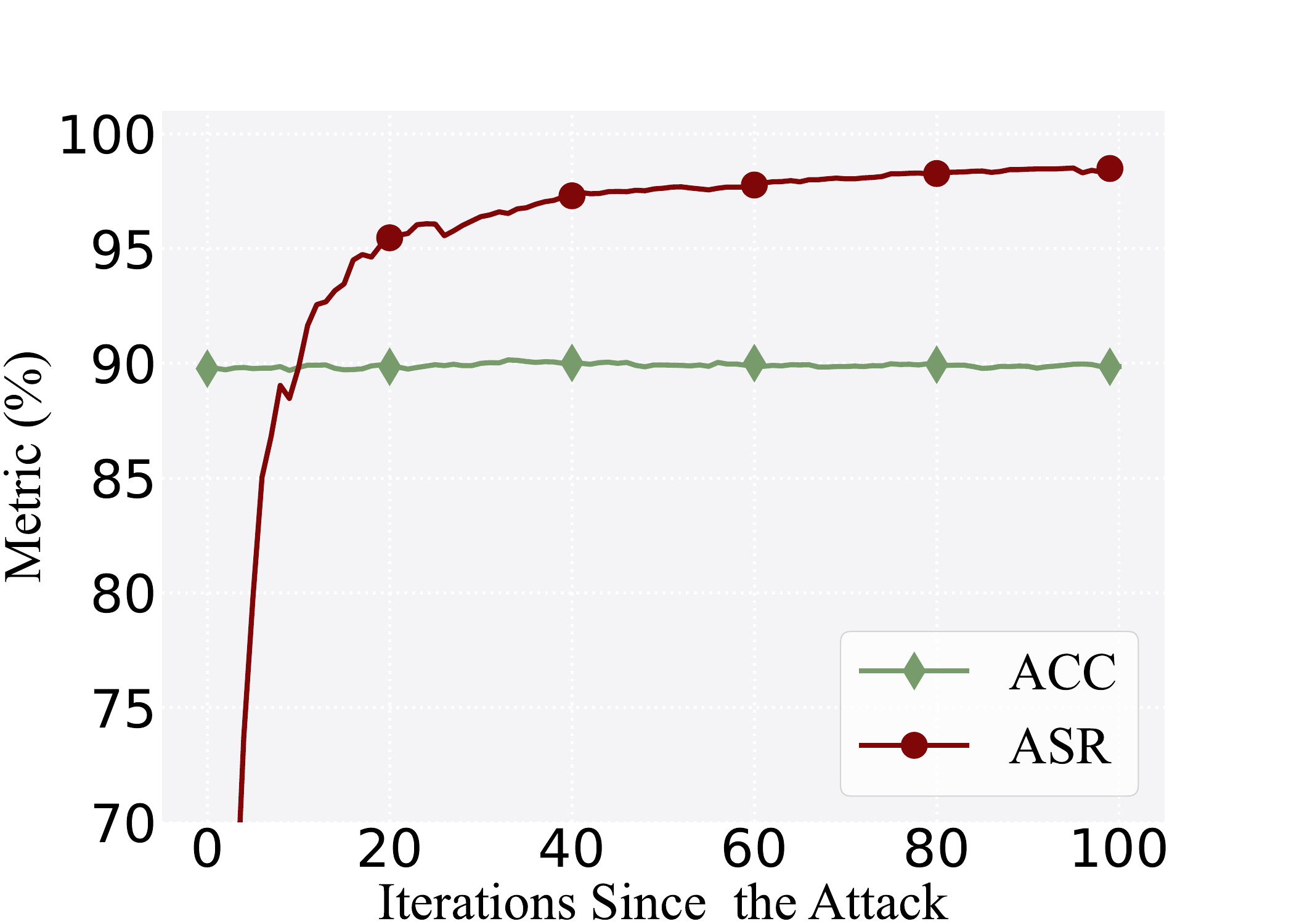}}
  \subfigure[RFLBAT]{\label{fig:lf_c}
		\includegraphics[width=0.38\columnwidth]{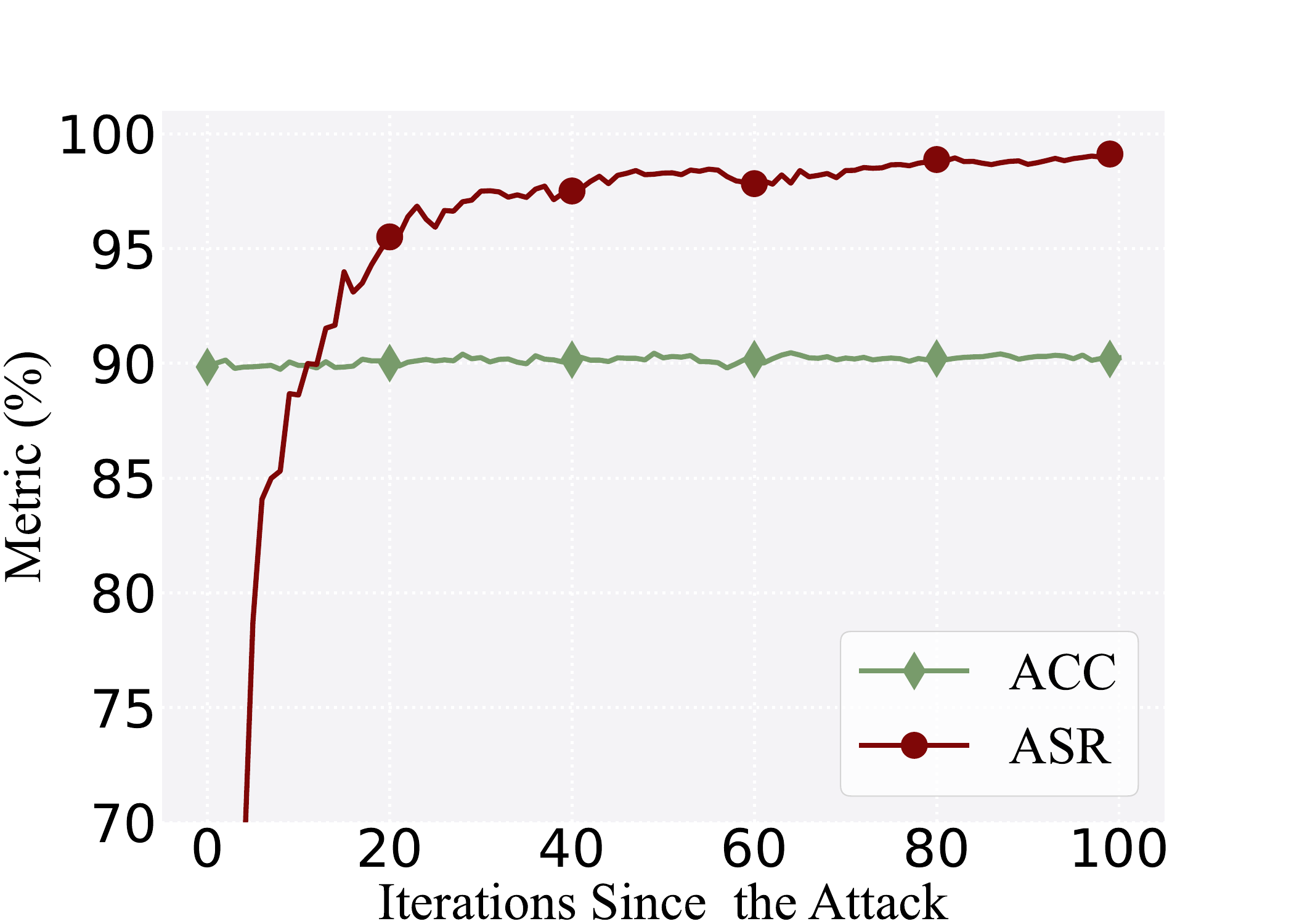}}
  \subfigure[FoolsGold]{\label{fig:lf_c}
		\includegraphics[width=0.38\columnwidth]{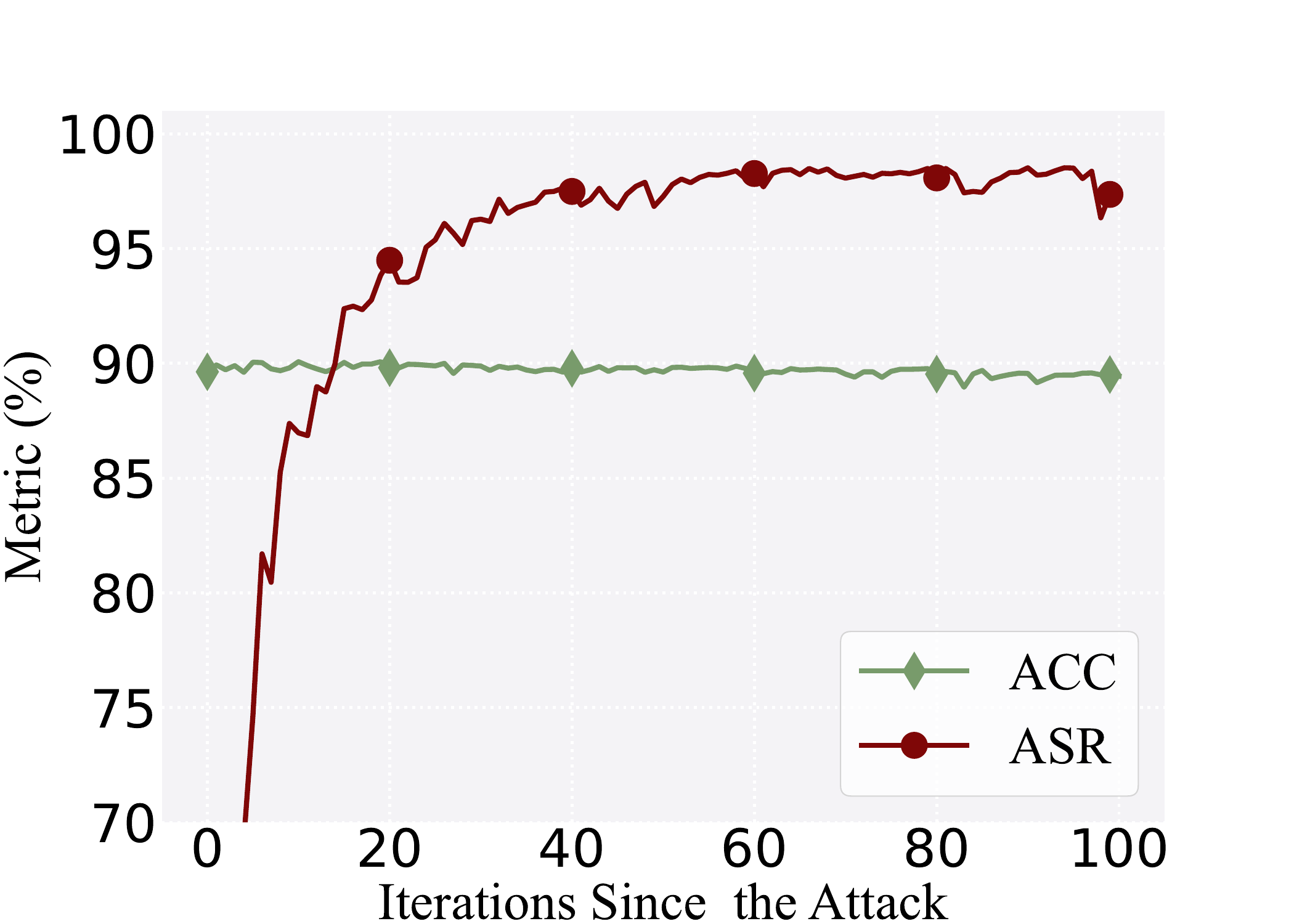}}
  \subfigure[FedAvg]{
		\includegraphics[width=0.38\columnwidth]{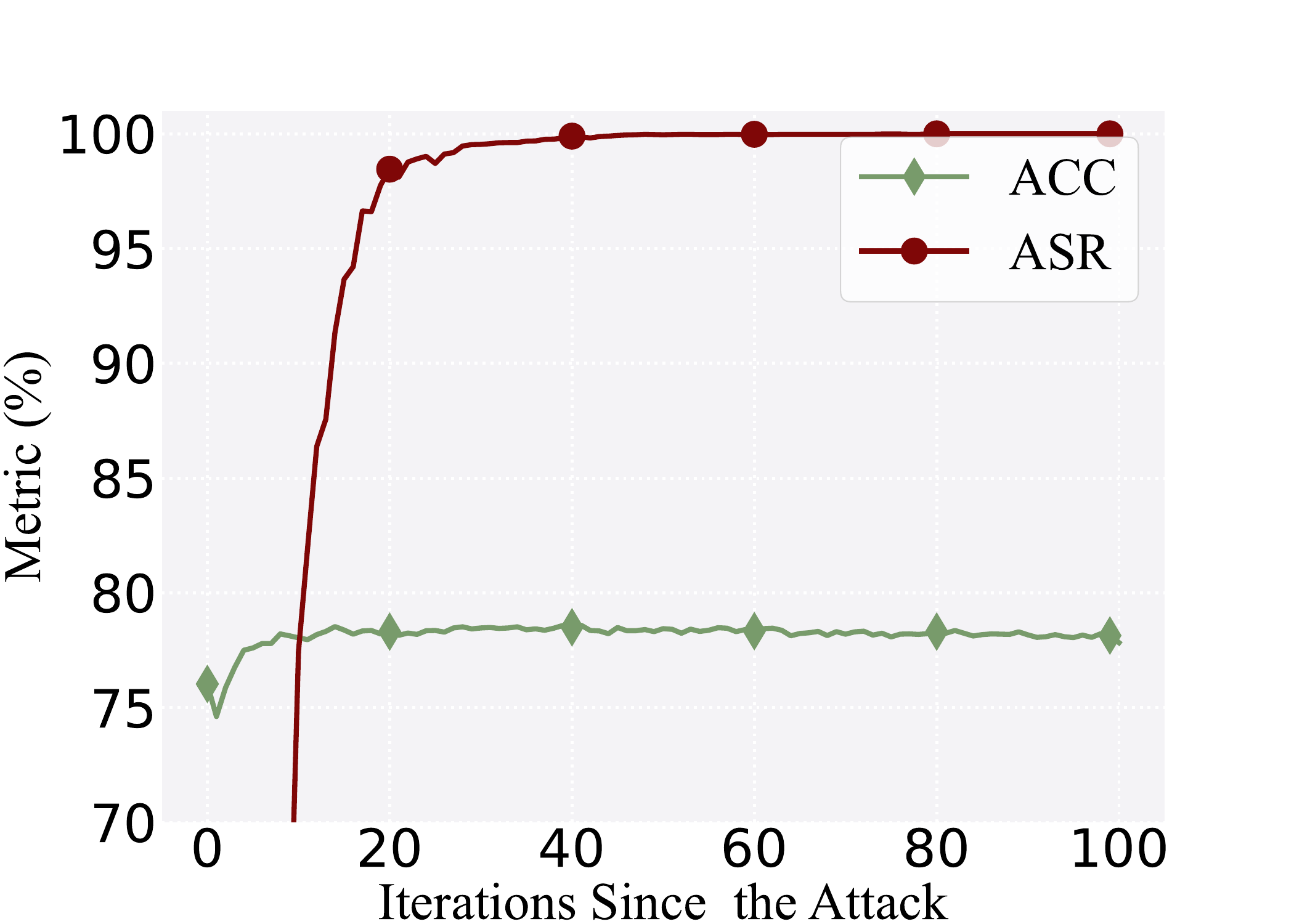}}
	\subfigure[Norm Clipping]{\label{fig:lf_c}
		\includegraphics[width=0.38\columnwidth]{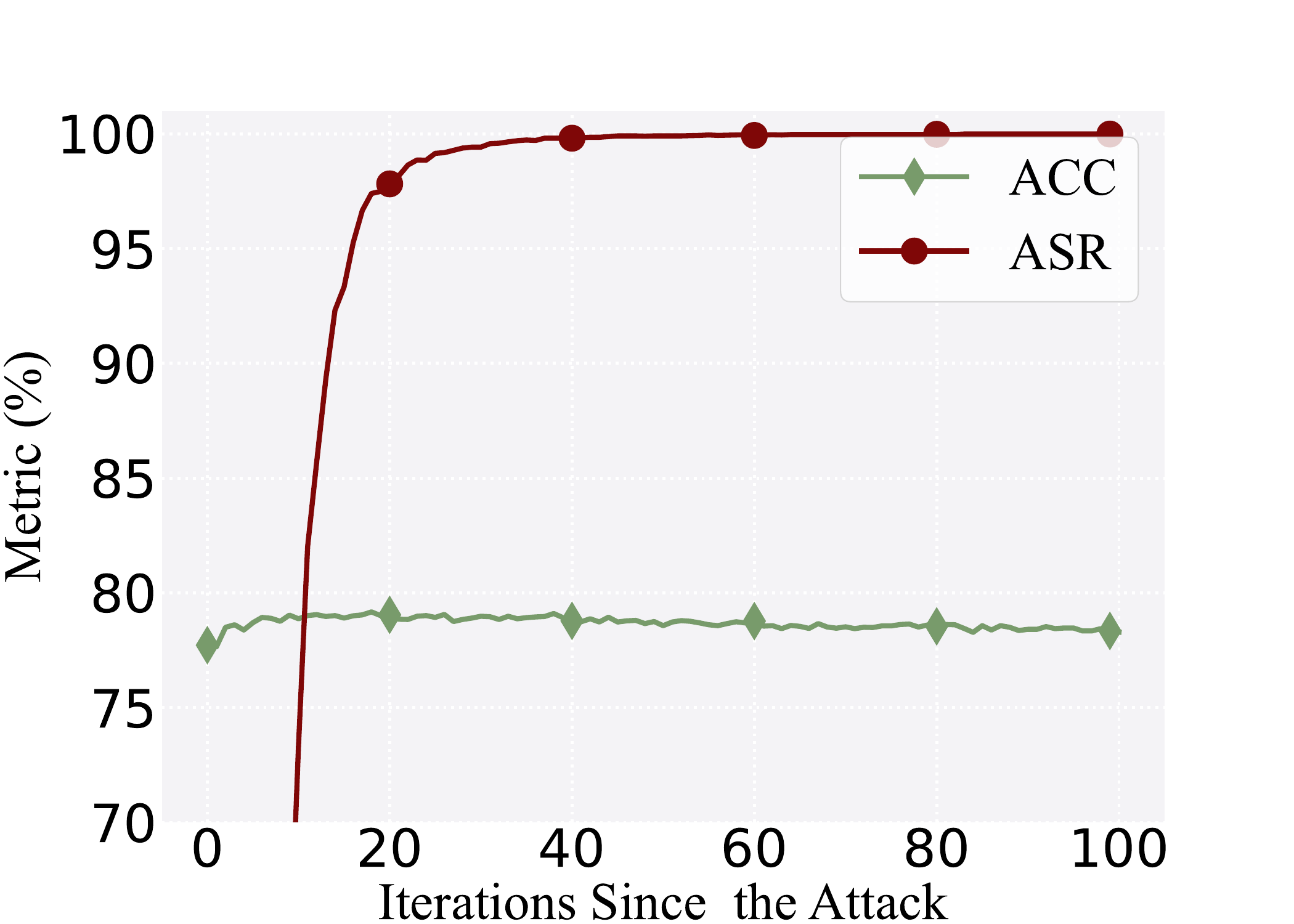}}
  \subfigure[FLAME]{\label{fig:lf_c}
		\includegraphics[width=0.38\columnwidth]{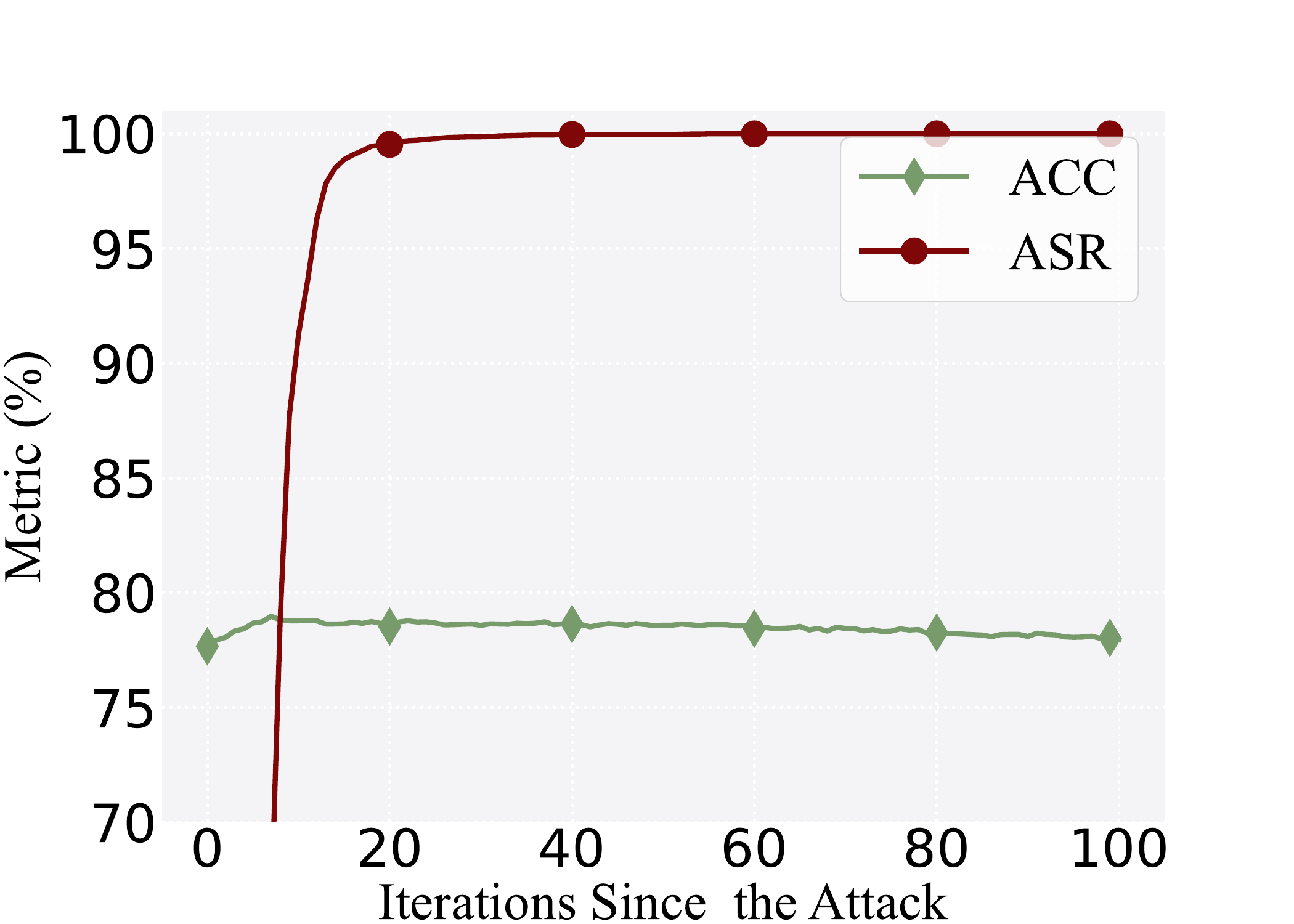}}
  \subfigure[RFLBAT]{\label{fig:lf_c}
		\includegraphics[width=0.38\columnwidth]{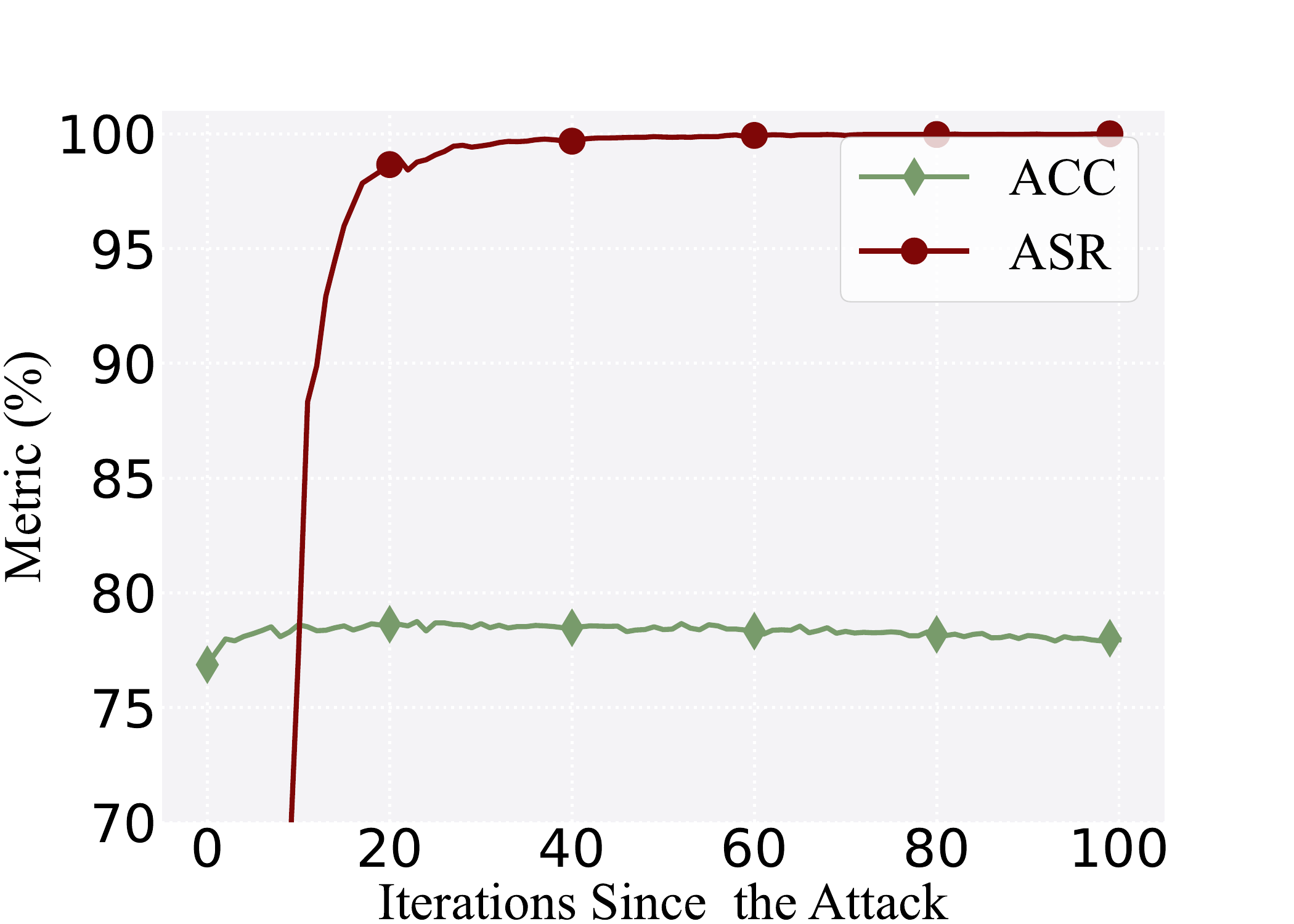}}
  \subfigure[FoolsGold]{\label{fig:lf_c}
		\includegraphics[width=0.38\columnwidth]{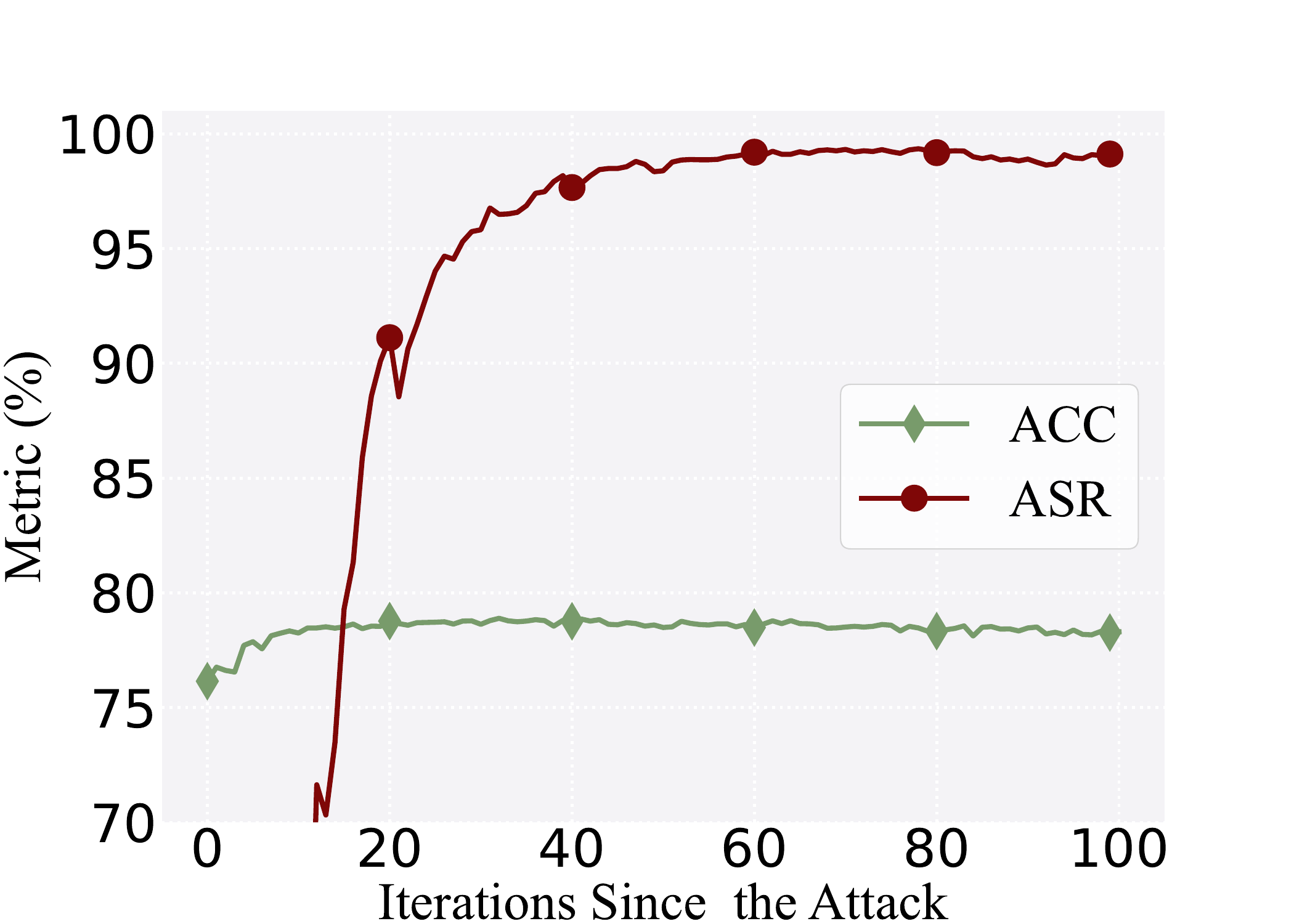}}
  \subfigure[FedAvg]{
		\includegraphics[width=0.38\columnwidth]{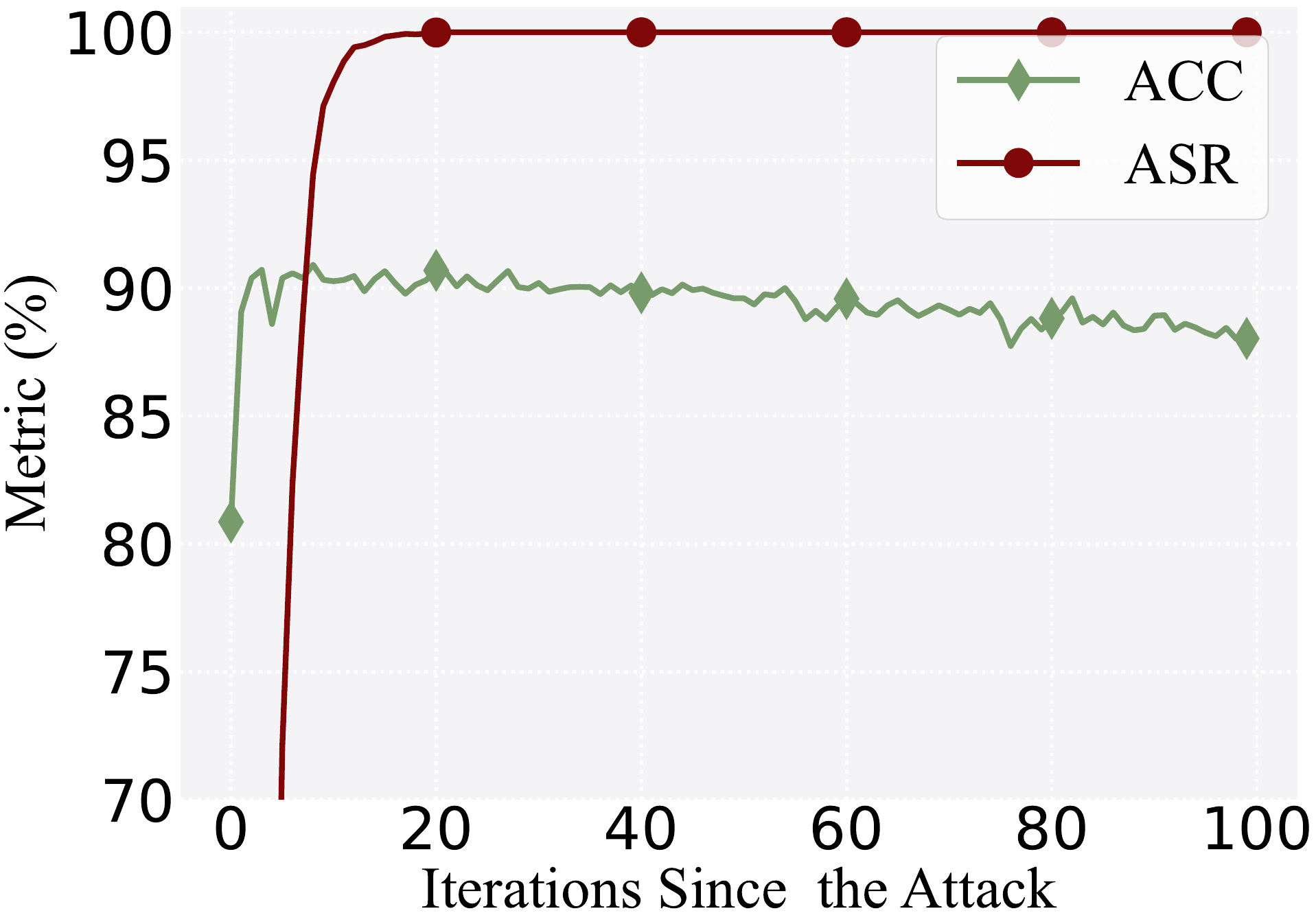}}
	\subfigure[Norm Clipping]{\label{fig:lf_c}
		\includegraphics[width=0.38\columnwidth]{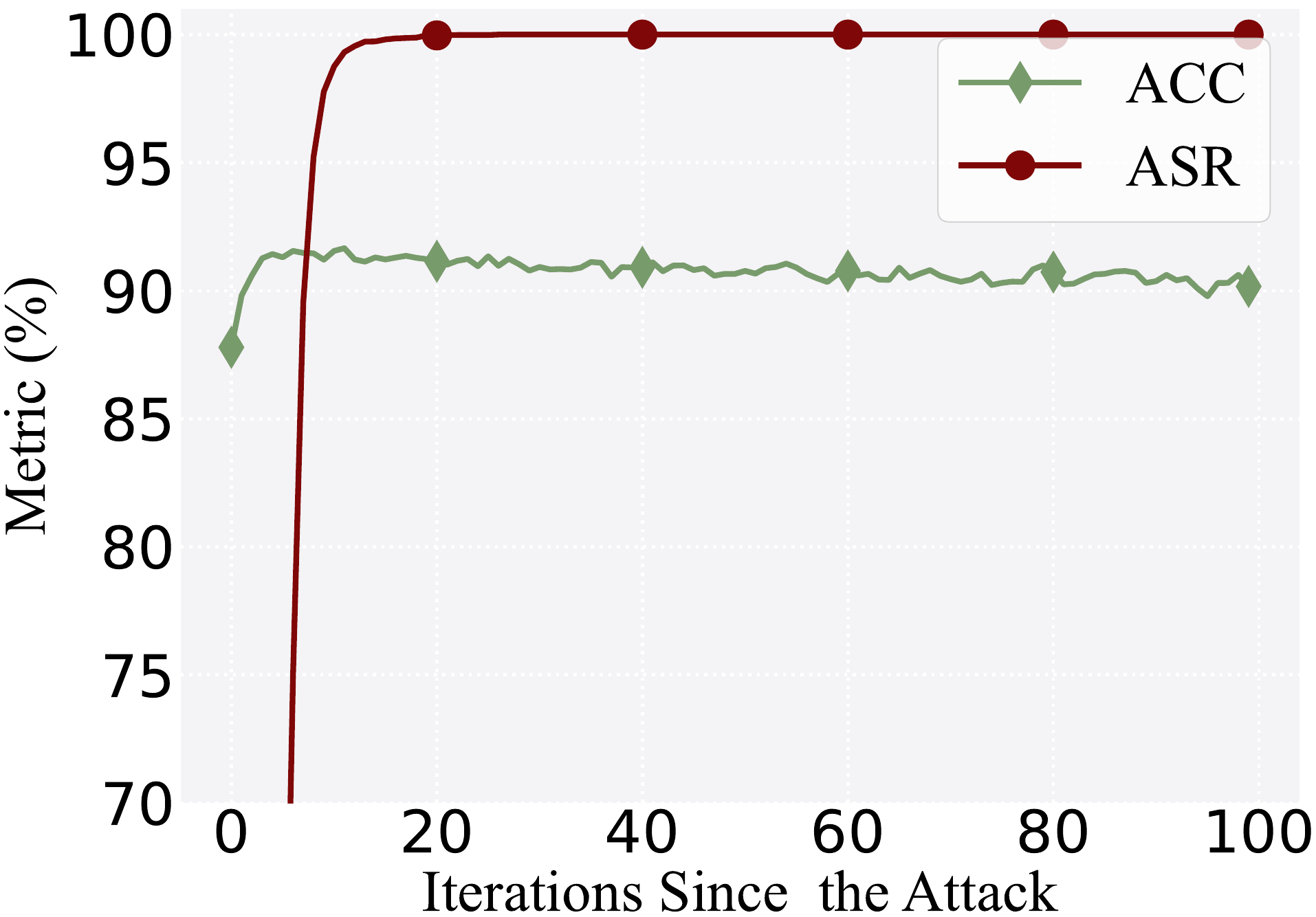}}
  \subfigure[FLAME]{\label{fig:lf_c}
		\includegraphics[width=0.38\columnwidth]{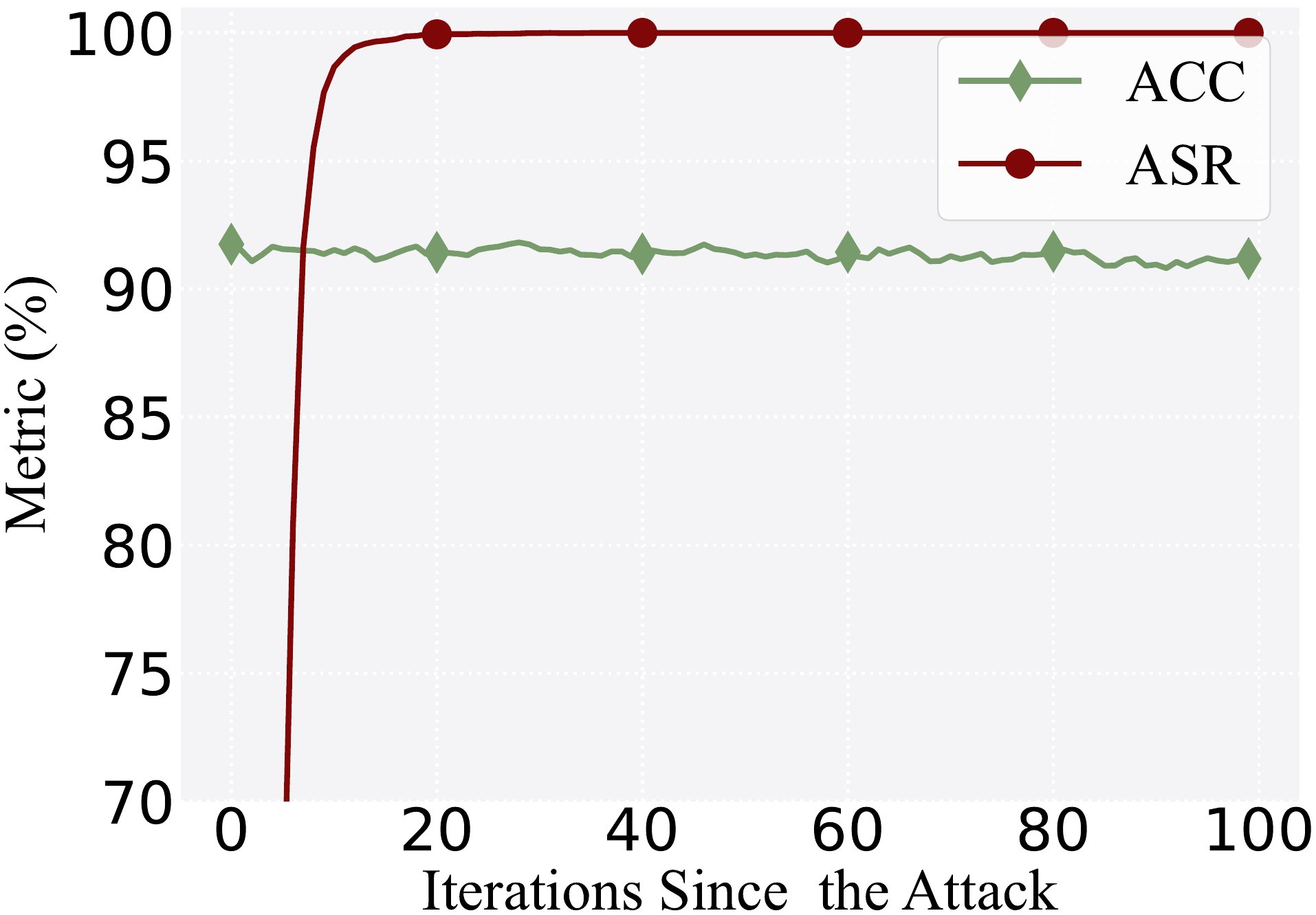}}
  \subfigure[RFLBAT]{\label{fig:lf_c}
		\includegraphics[width=0.38\columnwidth]{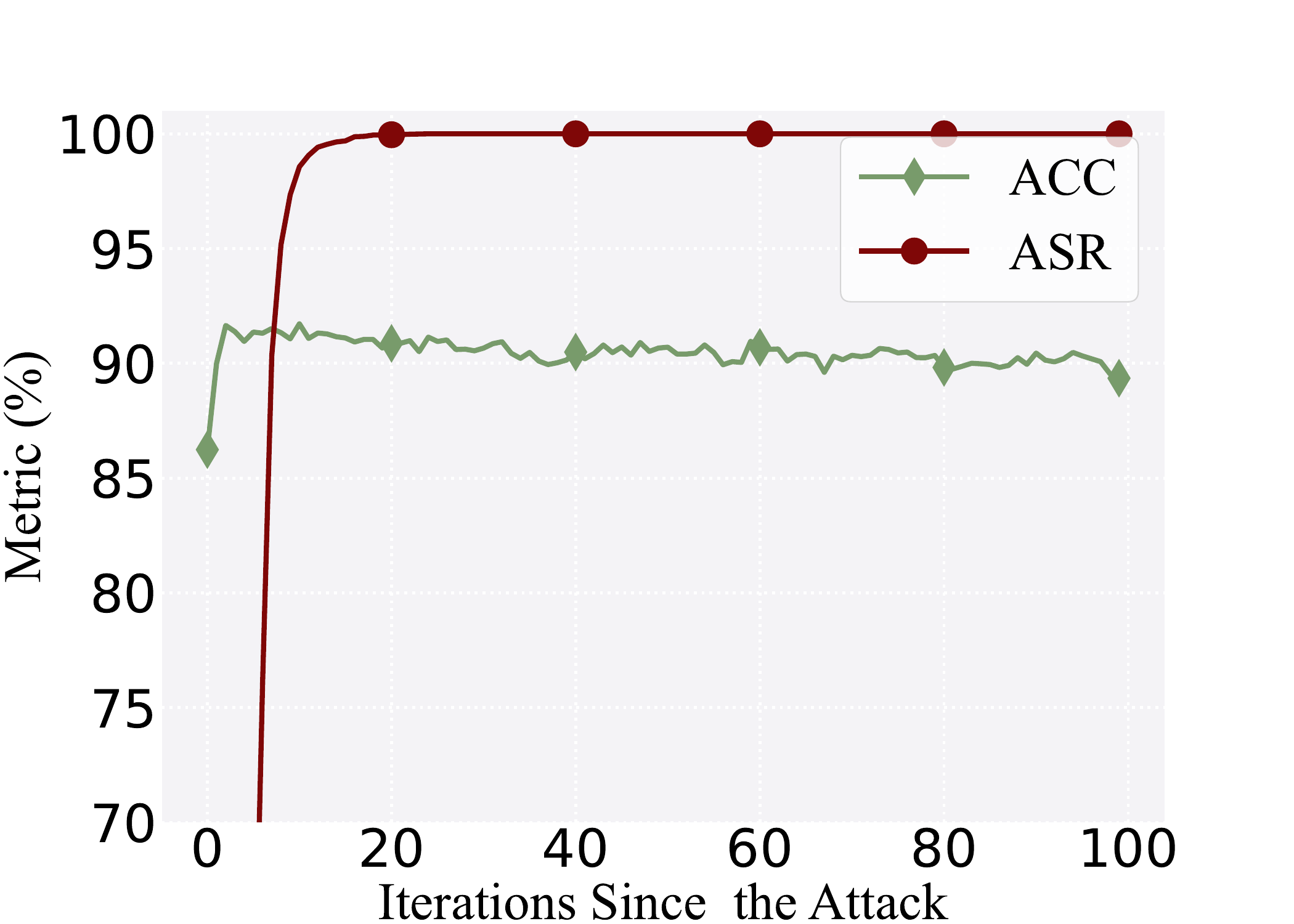}}
  \subfigure[FoolsGold]{\label{fig:lf_c}
		\includegraphics[width=0.38\columnwidth]{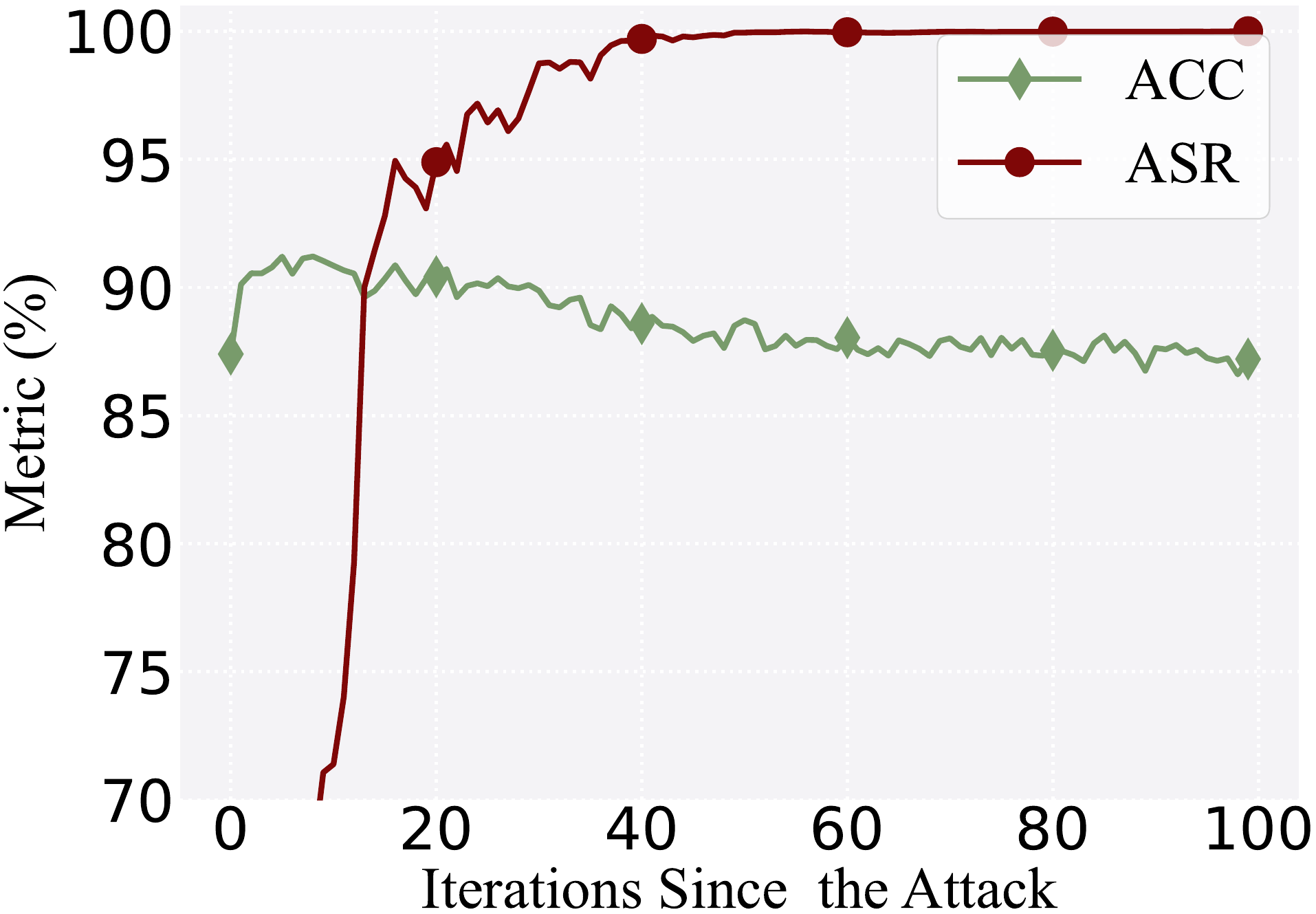}}
    \vspace{-1.5mm}
	\caption{Attack performance on CIFAR-10 (first row), CIFAR-100 (second row), and GTSRB (third row).}
  \vspace{-1.5mm}
	\label{fig:performance}
\end{figure*}

\begin{figure*}[t]
	\centering
	\subfigure[FedAvg]{
		\includegraphics[width=0.38\columnwidth]{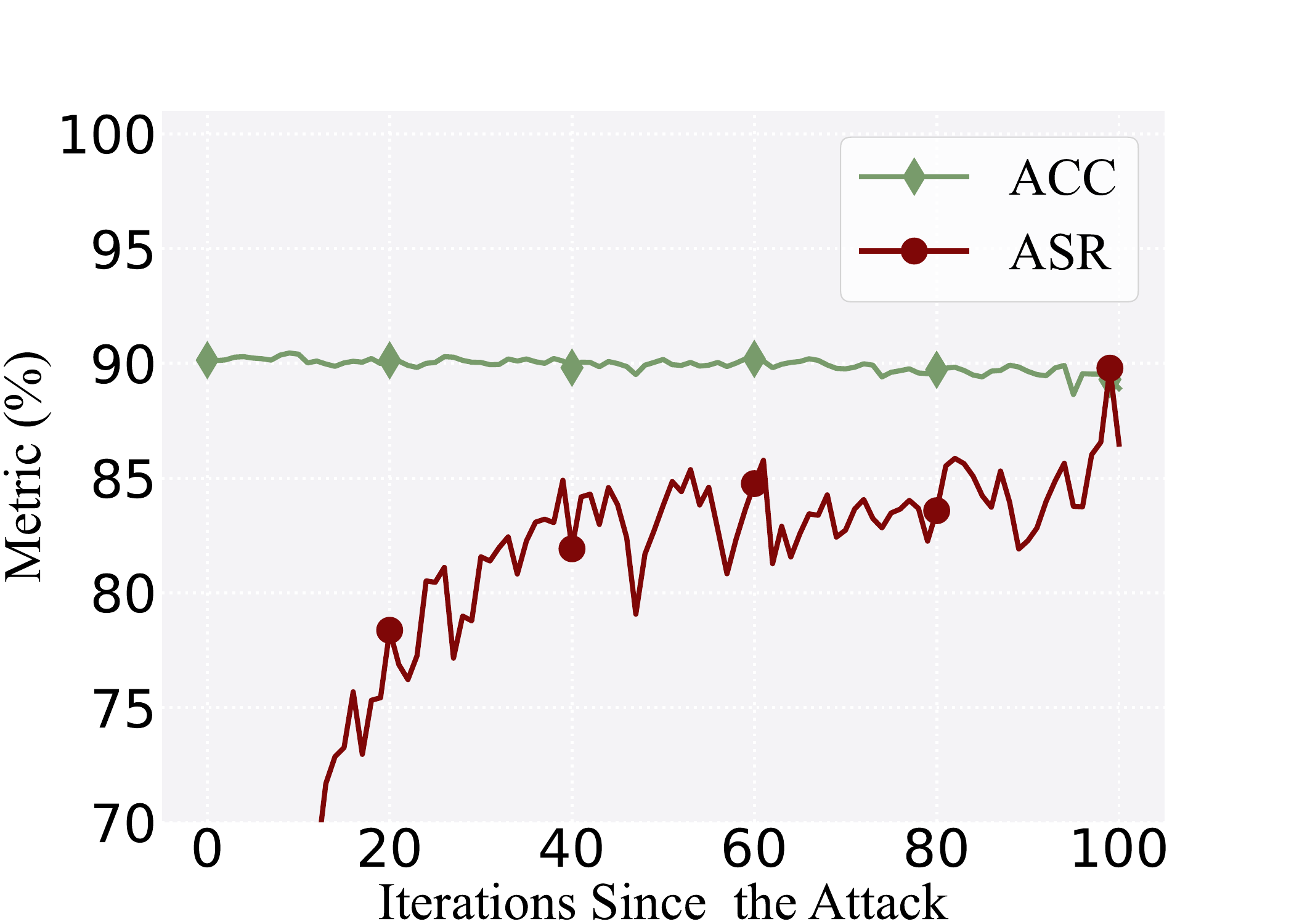}}
	\subfigure[Norm Clipping]{\label{fig:lf_c}
		\includegraphics[width=0.38\columnwidth]{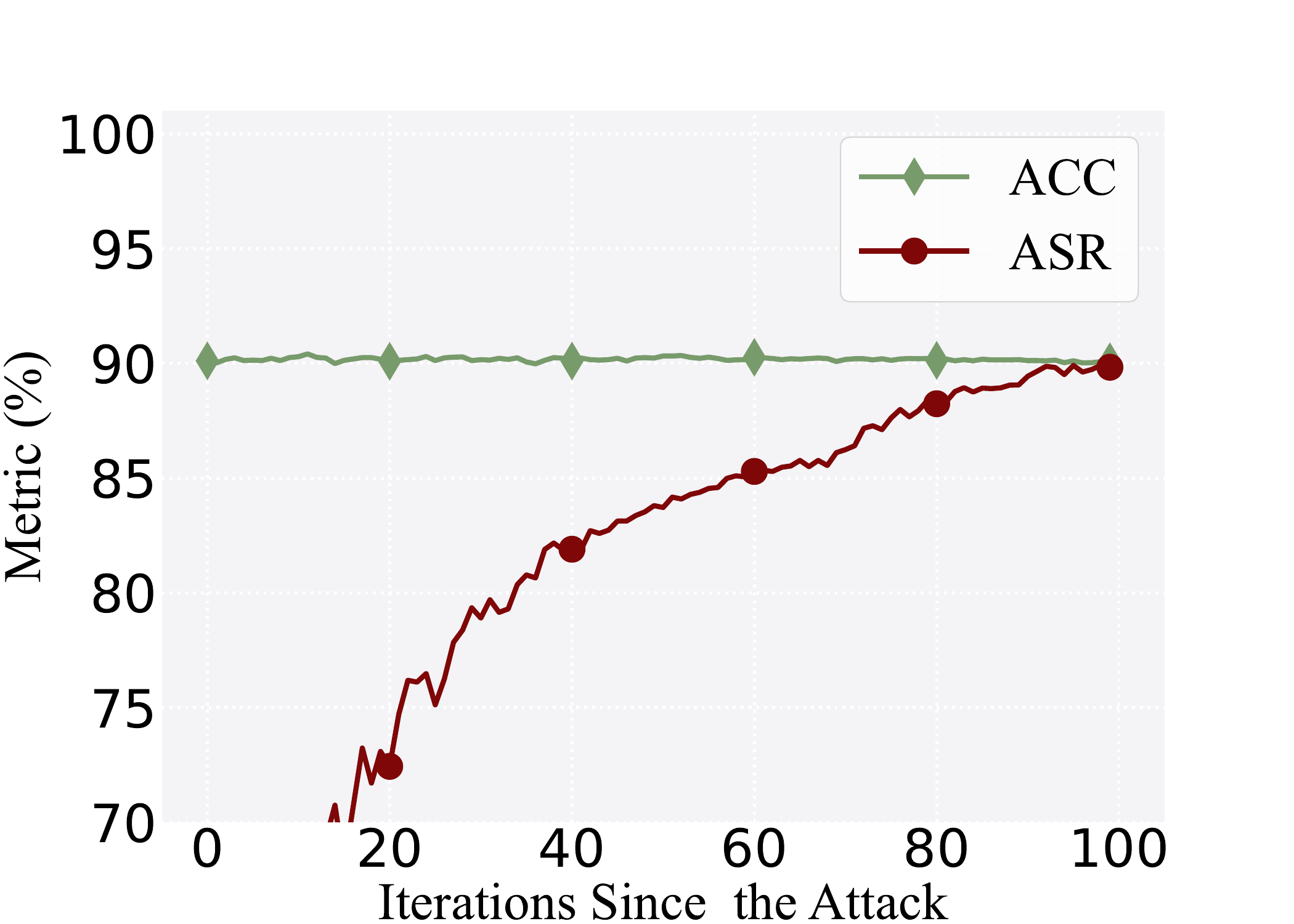}}
  \subfigure[FLAME]{\label{fig:lf_c}
		\includegraphics[width=0.38\columnwidth]{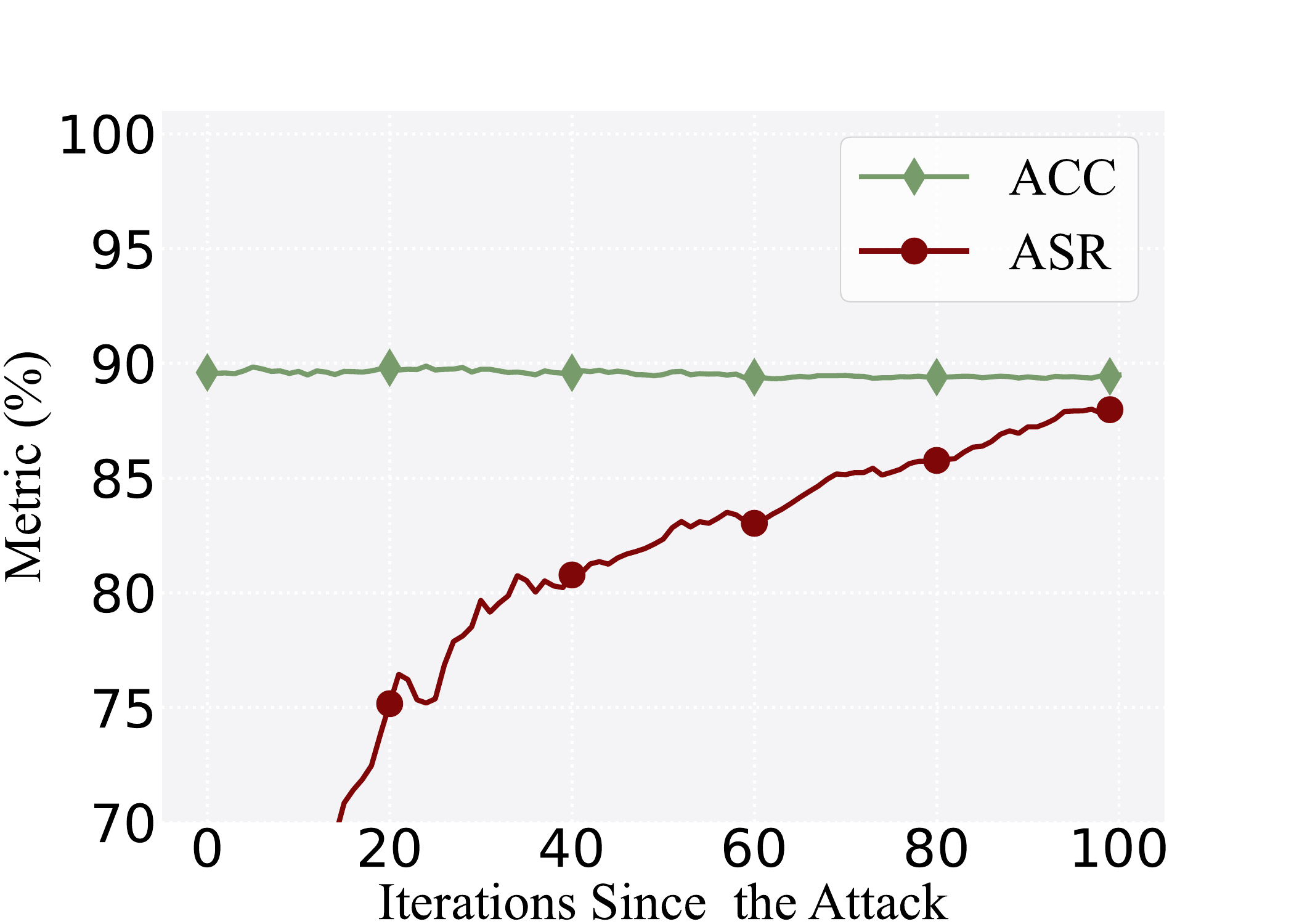}}
  \subfigure[RFLBAT]{\label{fig:lf_c}
		\includegraphics[width=0.38\columnwidth]{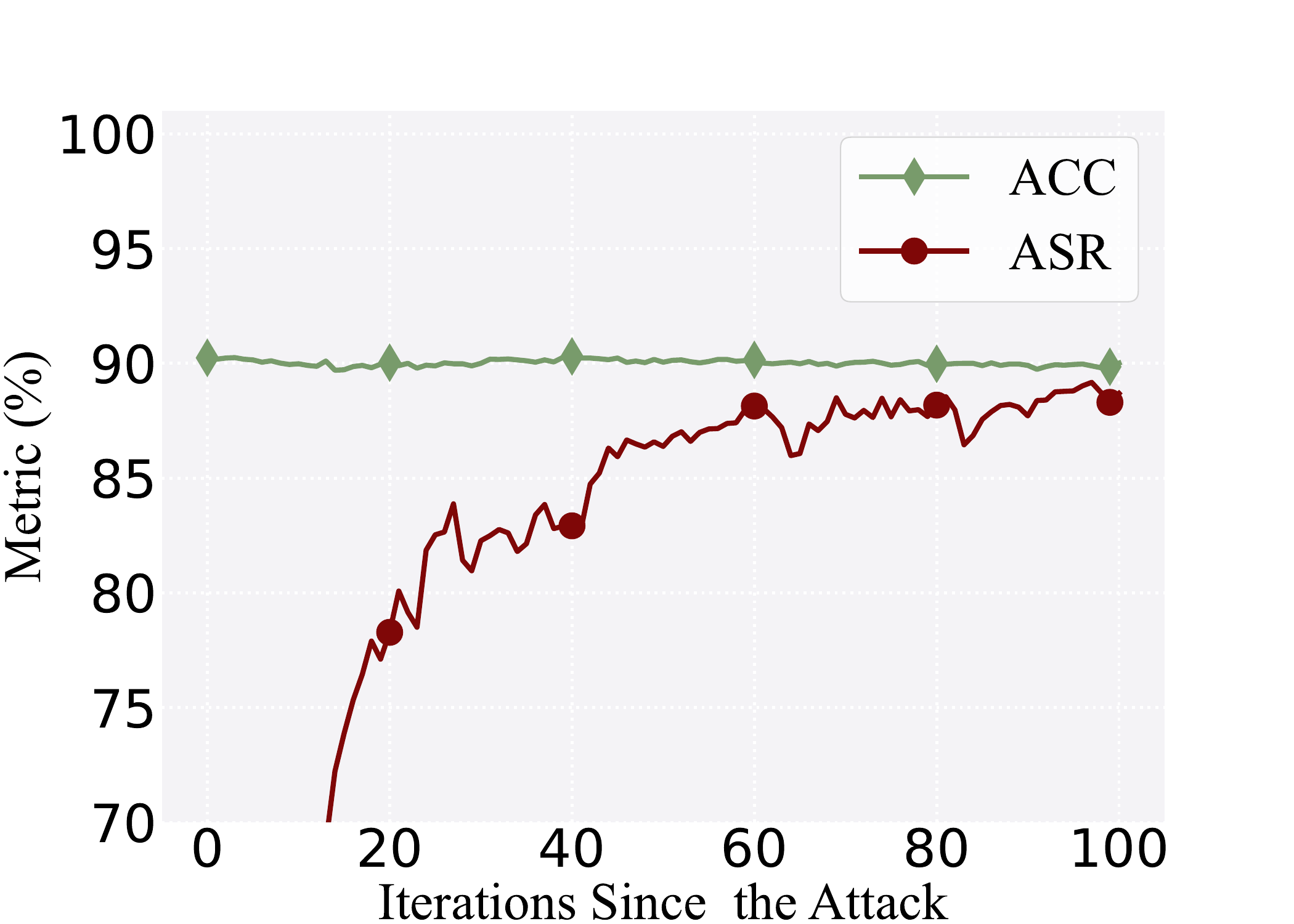}}
  \subfigure[FoolsGold]{\label{fig:lf_c}
		\includegraphics[width=0.38\columnwidth]{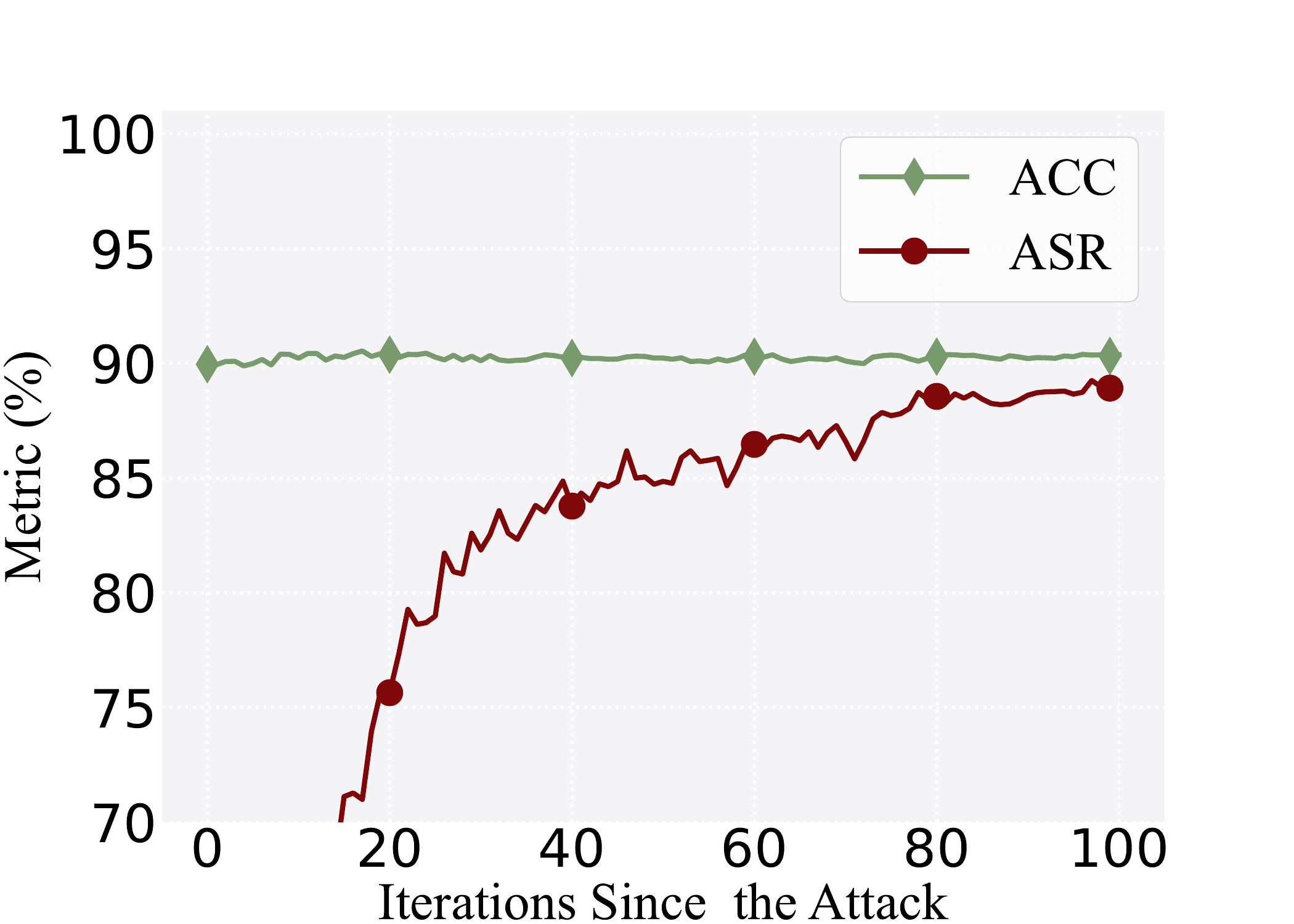}}
    \vspace{-1.5mm}
	\caption{Attack performance on CIFAR-10 with synthetic dataset.}
  \vspace{-1.5mm}
	\label{fig:self-constructed dataset performance}
\end{figure*}

% \begin{figure*}[t]
% 	\centering
% 	\subfigure[CIFAR100 FedAvg]{
% 		\includegraphics[width=0.38\columnwidth]{figs/cifar100_FedAvg.pdf}}
% 	\subfigure[Norm Clipping]{\label{fig:lf_c}
% 		\includegraphics[width=0.38\columnwidth]{figs/cifar100_normclipping.pdf}}
%   \subfigure[CIFAR100 FLAME]{\label{fig:lf_c}
% 		\includegraphics[width=0.38\columnwidth]{figs/cifar100_flame.pdf}}
%   \subfigure[CIFAR100 RFLBAT]{\label{fig:lf_c}
% 		\includegraphics[width=0.38\columnwidth]{figs/cifar100_RFLBAT.pdf}}
%   \subfigure[CIFAR100 FoolsGold]{\label{fig:lf_c}
% 		\includegraphics[width=0.38\columnwidth]{figs/cifar100_foolsgold.pdf}}
% 	\caption{Attack performance on CIFAR-100}
%  % \vspace{-5mm}
% 	\label{fig:performance_cifar100}
% \end{figure*}

% \begin{figure*}[t]
% 	\centering
% 	\subfigure[FedAvg]{
% 		\includegraphics[width=0.38\columnwidth]{figs/gtsrb_FedAvg.pdf}}
% 	\subfigure[Norm Clipping]{\label{fig:lf_c}
% 		\includegraphics[width=0.38\columnwidth]{figs/gtsrb_normclipping.pdf}}
%   \subfigure[FLAME]{\label{fig:lf_c}
% 		\includegraphics[width=0.38\columnwidth]{figs/gtsrb_flame.pdf}}
%   \subfigure[RFLBAT]{\label{fig:lf_c}
% 		\includegraphics[width=0.38\columnwidth]{figs/gtsrb_RFLBAT.pdf}}
%   \subfigure[FoolsGold]{\label{fig:lf_c}
% 		\includegraphics[width=0.38\columnwidth]{figs/gtsrb_foolsgold.pdf}}
% 	\caption{Attack performance on GTSRB}
%  % \vspace{-5mm}
% 	\label{fig:performance_gtsrb}
% \end{figure*}

\noindent\textbf{Evaluated defenses.}
We evaluate DarkFed against five SOTA defenses: FedAvg~\cite{FedAvg}, Norm Clipping~\cite{HowToBackdoor}, FLAME~\cite{FLAME}, RFLBAT~\cite{RFLBAT}, and FoolsGold~\cite{Sybils}. The defenses cover all the types outlined in Sec.~\ref{sec:backdoor defenses}, showcasing the universality of DarkFed.

\subsection{Experimental Results}
\noindent\textbf{Attack performance. }
We systematically verify the attainment of the adversary's goals to assess the attack performance. Figs.~\ref{fig:performance} illustrates the impact of DarkFed on SOTA defenses on CIFAR-10 (first row), CIFAR-100 (second row), and GTSRB (third row). The green line represents ACC, while the red line denotes ASR. In terms of fidelity, DarkFed achieves high ACC on both CIFAR-10 and CIFAR-100, with an accuracy degradation within $1\%$ compared to the initial model. On GTSRB, across various defenses, accuracy degradation ranges between $1\%$ and $3\%$. These marginal accuracy degradations do not significantly affect model usability, showcasing DarkFed's fidelity achievement. In terms of effectiveness, DarkFed rapidly achieves nearly $100\%$ ASR with few iterations across all three datasets (around 20 iterations for CIFAR-10 and CIFAR-100, and 10 iterations for GTSRB), highlighting its remarkable attack effectiveness. The stealthiness goal is indirectly demonstrated through the preceding two goals, as a sufficiently concealed attack is essential to ensure that backdoor updates evade defenses, ultimately resulting in a global model that excels in both the main and backdoor tasks.

\noindent\textbf{Impact of attacker ratio. }
Tab.\ref{tab: impact of attacker percentage} illustrates the impact of attacker ratio on DarkFed under FLAME. It is noticeable that as the ratio of attackers increases, ACC experiences a slight decrease, while ASR exhibits an upward trend. However, overall, DarkFed shows minimal susceptibility to changes in the attacker ratio. Even in the presence of a $5\%$ attacker ratio, the ASR remains notably high, reaching a minimum of $94.30\%$. The ASR nearly peaks when the attacker ratio reaches $10\%$.

\noindent\textbf{Comparison with SOTA attacks. }
Existing research has not delved into data-free backdoor attacks in FL, thus we showcase DarkFed's superiority by directly comparing it with SOTA data-dependent attacks. Specifically, we consider the classic Model Replacement Attack~\cite{HowToBackdoor} and the latest 3DFed~\cite{3DFed}. It's important to note that these two attacks directly utilize task-specific data, while DarkFed relies solely on shadow dataset. Tab.~\ref{tab: Comparison with SOTA attacks} presents the comparative results on CIFAR-10.
In terms of ACC, these three attacks exhibit no significant differences; all maintain high model accuracy. On average, the differences among them are within $0.11\%$. Regarding ASR, Model Replacement Attack performs relatively poorly, only managing to backdoor FedAvg and Norm Clipping. Both 3DFed and DarkFed can overcome all defenses, but DarkFed's average ASR is slightly higher than 3DFed. Furthermore, we observe that 3DFed's ASR under the RFLBAT defense is $3\%$ lower than DarkFed. We speculate that this is because 3DFed needs to use decoy model updates as bait to mislead RFLBAT. As a result, not all backdoor updates can be accepted by RFLBAT, sacrificing attack performance to some extent.

\begin{table}[!t]
\renewcommand\arraystretch{1}
\centering
% \vspace{-4mm}
\caption{Impact of the attacker ratio.}
  \vspace{-1.5mm}
\resizebox{\linewidth}{!}{
\begin{tabular}{|c||cc|cc|cc|}
\hline
\multirow{2}{*}{\textbf{\begin{tabular}[c]{@{}c@{}}Attacker   \\ Ratio\end{tabular}}} &
  \multicolumn{2}{c|}{\textbf{CIFAR-10}} &
  \multicolumn{2}{c|}{\textbf{CIFAR-100}} &
  \multicolumn{2}{c|}{\textbf{GTSRB}} \\ \cline{2-7} 
 &
  \multicolumn{1}{c|}{\textbf{ACC (\%)}} &
  \textbf{ASR (\%)} &
  \multicolumn{1}{c|}{\textbf{ACC (\%)}} &
  \textbf{ASR (\%)} &
  \multicolumn{1}{c|}{\textbf{ACC (\%)}} &
  \textbf{ASR (\%)} \\ \hline
5\%  & \multicolumn{1}{c|}{90.61} & 95.85 & \multicolumn{1}{c|}{79.13} & 94.30  & \multicolumn{1}{c|}{92.82} & 99.94  \\ \hline
10\% & \multicolumn{1}{c|}{90.22} & 97.81 & \multicolumn{1}{c|}{78.94} & 99.77  & \multicolumn{1}{c|}{92.53} & 100.00 \\ \hline
15\% & \multicolumn{1}{c|}{90.13} & 98.51 & \multicolumn{1}{c|}{78.82} & 99.92  & \multicolumn{1}{c|}{91.84} & 100.00 \\ \hline
20\% & \multicolumn{1}{c|}{90.04} & 98.96 & \multicolumn{1}{c|}{78.62} & 100.00 & \multicolumn{1}{c|}{91.74} & 100.00 \\ \hline
25\% & \multicolumn{1}{c|}{90.09} & 99.01 & \multicolumn{1}{c|}{78.15} & 100.00 & \multicolumn{1}{c|}{91.51} & 100.00 \\ \hline
\end{tabular}
}
\vspace{-3mm}
\label{tab: impact of attacker percentage}
\end{table}

\noindent\textbf{Attack with synthetic dataset. }The preceding experiments utilize shadow datasets comprising real data from vastly different domains, yielding highly effective attack outcomes. Consequently, a naturally intriguing question arises: Can we achieve similar attack results with synthetic dataset? To answer this question, we employ Gauss-I as the shadow dataset for CIFAR-10, and the experimental results are depicted in Fig.~\ref{fig:self-constructed dataset performance}. Compared with the earlier experiments (the first row of Fig.~\ref{fig:performance}), ACC remains consistent, hovering around $90\%$. Although ASR suffers a relative decrease, it still approaches $90\%$. The experimental results are promising, indicating that even without crawling any datasets online, the use of self-constructed, semantically meaningless data can successfully inject a backdoor without compromising model accuracy.

% \begin{table}[!t]
% \renewcommand\arraystretch{1}
% \centering
% % \vspace{-4mm}
% \caption{}
% % \vspace{-4mm}
% \resizebox{\linewidth}{!}{
% \begin{tabular}{|c|cc|cc|cc|}
% \hline
% \multirow{2}{*}{Defenses} & \multicolumn{2}{c|}{CIFAR-10}      & \multicolumn{2}{c|}{CIFAR-100}      & \multicolumn{2}{c|}{GTSRB}          \\ \cline{2-7} 
%                 & \multicolumn{1}{c|}{ACC(\%)} & ASR(\%) & \multicolumn{1}{c|}{ACC(\%)} & ASR(\%) & \multicolumn{1}{c|}{ACC(\%)} & ASR(\%) \\ \hline
% FedAvg                    & \multicolumn{1}{c|}{89.54} & 99.18 & \multicolumn{1}{c|}{78.36} & 100.00 & \multicolumn{1}{c|}{90.67} & 100.00 \\ \hline
% Norm   Clipping & \multicolumn{1}{c|}{90.26}   & 99.20   & \multicolumn{1}{c|}{78.57}   & 99.99   & \multicolumn{1}{c|}{91.12}   & 100.00  \\ \hline
% FLAME                     & \multicolumn{1}{c|}{89.96} & 98.51 & \multicolumn{1}{c|}{78.62} & 100.00 & \multicolumn{1}{c|}{91.74} & 100.00 \\ \hline
% RFLBAT                    & \multicolumn{1}{c|}{90.24} & 99.18 & \multicolumn{1}{c|}{78.02} & 100.00 & \multicolumn{1}{c|}{91.14} & 100.00 \\ \hline
% FoolsGold                 & \multicolumn{1}{c|}{89.49} & 98.52 & \multicolumn{1}{c|}{78.47} & 99.35  & \multicolumn{1}{c|}{89.27} & 99.45  \\ \hline
% \end{tabular}
% }
% % \vspace{-5mm}
% \label{tab: Patameter settings}
% \end{table}

\begin{table}[!t]
\renewcommand\arraystretch{1}
\centering
% \vspace{-4mm}
\caption{Comparison with SOTA data-dependent attacks.}
  \vspace{-1.5mm}
\resizebox{\linewidth}{!}{
\begin{tabular}{|c||cc|cc|cc|}
\hline
 &
  \multicolumn{2}{c|}{\textbf{Model Replacement}} &
  \multicolumn{2}{c|}{\textbf{3DFed}} &
  \multicolumn{2}{c|}{\textbf{DarkFed}} \\ \cline{2-7} 
\multirow{-2}{*}{\textbf{Defense}} &
  \multicolumn{1}{c|}{\textbf{ACC (\%)}} &
  \textbf{ASR (\%)} &
  \multicolumn{1}{c|}{\textbf{ACC (\%)}} &
  \textbf{ASR (\%)} &
  \multicolumn{1}{c|}{\textbf{ACC (\%)}} &
  \textbf{ASR (\%)} \\ \hline
FedAvg &
  \multicolumn{1}{c|}{{\color[HTML]{009901} \textbf{89.74}}} &
  99.16 &
  \multicolumn{1}{c|}{89.66} &
  {\color[HTML]{9A0000} \textbf{99.85}} &
  \multicolumn{1}{c|}{89.54} &
  99.18 \\ \hline
Norm   Clipping &
  \multicolumn{1}{c|}{90.07} &
  97.93 &
  \multicolumn{1}{c|}{90.14} &
  98.71 &
  \multicolumn{1}{c|}{{\color[HTML]{009901} \textbf{90.26}}} &
  {\color[HTML]{9A0000} \textbf{99.20}} \\ \hline
FLAME &
  \multicolumn{1}{c|}{{\color[HTML]{009901} \textbf{90.26}}} &
  9.74 &
  \multicolumn{1}{c|}{90.01} &
  {\color[HTML]{9A0000} \textbf{99.89}} &
  \multicolumn{1}{c|}{89.96} &
  98.51 \\ \hline
RFLBAT &
  \multicolumn{1}{c|}{90.17} &
  8.84 &
  \multicolumn{1}{c|}{90.21} &
  96.18 &
  \multicolumn{1}{c|}{{\color[HTML]{009901} \textbf{90.24}}} &
  {\color[HTML]{9A0000} \textbf{99.18}} \\ \hline
FoolsGold &
  \multicolumn{1}{c|}{89.82} &
  9.77 &
  \multicolumn{1}{c|}{{\color[HTML]{009901} \textbf{89.97}}} &
  98.51 &
  \multicolumn{1}{c|}{89.49} &
  {\color[HTML]{9A0000} \textbf{98.52}} \\ \hline
  Average &
  \multicolumn{1}{c|}{{\color[HTML]{009901} \textbf{90.01}}} &
  45.09 &
  \multicolumn{1}{c|}{90.00} &
  98.63 &
  \multicolumn{1}{c|}{89.90} &
  {\color[HTML]{9A0000} \textbf{98.92}} \\ \hline
\end{tabular}
}
\vspace{-3mm}
\label{tab: Comparison with SOTA attacks}
\end{table}

% \section{Limitations}
\vspace{-1.5mm}
\section{Conclusion}
This paper introduces DarkFed, the inaugural data-free backdoor attack in FL. DarkFed eliminates the reliance on main task-related data, rendering it suitable for fake client scenarios and offering an effective solution for practical backdoor attacks. To heighten the attack's stealthiness, we propose property mimicry, mimicking benign updates' properties to confound defenses. Extensive experiments demonstrate that DarkFed attains performance comparable to SOTA data-dependent attacks.

\section*{Acknowledgments}
Minghui’s work is supported in part by the National Natural Science Foundation of China (Grant No. 62202186). Shengshan’s work is supported in part by the National Natural Science Foundation of China (Grant Nos. 62372196, U20A20177).

% This work is supported by the National Natural Science Foundation of China (Grant No.U20A20177) and Hubei Province Key R\&D Technology Special Innovation Project (Grant No.2021BAA032).
%% The file named.bst is a bibliography style file for BibTeX 0.99c
\bibliographystyle{named}
\bibliography{ijcai24}

\end{document}